\documentclass[superscriptaddress,prx,twocolumn,longbibliography]{revtex4-1}
\usepackage{graphicx}
\usepackage{amsmath,amssymb,physics,dsfont}
\usepackage{color}
\usepackage{hyperref}
\usepackage{microtype}
\usepackage{braket}
\usepackage{multirow}
\usepackage{array} % Add this in the preamble if not already present
\usepackage{booktabs}
\usepackage{comment}
\usepackage{makecell}

\newcommand{\tocentry}[2]{%
  \noindent\hyperref[#1]{\textcolor{blue}{\ref*{#1}.~#2}}\dotfill
  \hyperref[#1]{\textcolor{blue}{\pageref*{#1}}}\par}

\newcommand{\tocsubentry}[2]{%
  \noindent\hspace*{1.5em}\hyperref[#1]{\textcolor{blue}{\ref*{#1}.~#2}}\dotfill
  \hyperref[#1]{\textcolor{blue}{\pageref*{#1}}}\par}

\begin{document}

\title{CDW Gap Collapse and Weyl State Restoration in \texorpdfstring{Ta$_2$Se$_8$I}{Ta2Se8I} via Coherent Phonons:\\
A First-Principles Study}

\author{Tao Jiang}
\email{tao.jiang@rice.edu}
\affiliation{Ames National Laboratory, U.S. Department of Energy, Ames, Iowa 50011, USA}
\affiliation{Department of Materials Science and Nanoengineering, Rice University, Houston, Texas 77005, USA}
\author{Jigang Wang}
\affiliation{Ames National Laboratory, U.S. Department of Energy, Ames, Iowa 50011, USA}
\affiliation{Department of Physics and Astronomy, Iowa State University, Ames, Iowa 50011, USA}

\author{Yong-Xin Yao}
\affiliation{Ames National Laboratory, U.S. Department of Energy, Ames, Iowa 50011, USA}
\affiliation{Department of Physics and Astronomy, Iowa State University, Ames, Iowa 50011, USA}

\begin{abstract}
\begin{center}
\textbf{Abstract}
\end{center}
Coherent phonon excitation offers a nonthermal route to control quantum phases of condensed matter. In this work, we employ first-principles calculations to investigate the phonon landscape of (TaSe$_4$)$_2$I in its charge-density-wave (CDW) phase. We identify nine symmetry-preserving Raman-active modes that can suppress the $\Gamma\text{--}\mathrm{Z}$ direct gap to the meV scale and render the system globally gapless by generating Weyl nodes at generic \(k\) points. Among them, the 2.52~THz CDW amplitude mode \(A(18)\) directly weakens the Ta-chain tetramerization, approaching a transient restoration of the uniform-chain geometry, while its low frequency and relatively small, Ta-dominated displacement make it the most efficient mode. Other Raman modes, dominated by Se vibrations, require significantly larger displacements to reach the Weyl-semimetallic regime and are generally less effective than \(A(18)\) at reducing the Ta-chain tetramerization. Furthermore, among the IR-active modes considered, the low-frequency \(B_3(7)\) mode (1.15~THz) provides the most favorable channel for nonlinear modulation of the \(A(18)\) potential. Although the coupling favors displacement toward the CDW-suppressing direction, the estimated indirect response remains far below that required for strong gap suppression. Direct excitation of \(A(18)\) therefore remains the most effective pathway to the Weyl-semimetallic regime.
\end{abstract}

\maketitle

\section*{Introduction}
Charge density wave (CDW), characterized by a periodic modulation of the electronic density often coupled to the underlying lattice, can emerge in low-dimensional materials where reduced dimensionality enhances electron-phonon and electron-electron interactions~\cite{wilson1975charge,gruner2018density,rossnagel2011origin,monceau2012electronic}. In quasi-one-dimensional compounds like (TaSe$_4$)$_2$I, this collective ordering leads to a metal-insulator transition below 263 K, stabilized via Peierls-like lattice distortions~\cite{wang1983charge,fujishita1985x,forro1987hall,lorenzo1998neutron,nemetha2001nmr,tournier2013electronic}. The CDW phase exhibits rich emergent phenomena that manifest in nonlinear transport, tunable resistivity, and strong coherent lattice responses, which underpin diverse applications in electronic devices~\cite{balandin2022one,ma2024memristive}, low-dissipation electron transport~\cite{ghosh2023low}, photodetection~\cite{cheng2021high,xiao2025light}, THz emission~\cite{kim2023observation}, photo-elasticity~\cite{cheng2023study} and phase change functionality~\cite{vaskivskyi2015controlling,song2023ultrafast}, as well as emerging opportunities in nonlinear optics, such as high-order harmonic generation~\cite{mckay2021optical,duncan2025coupled,hu2026charge}.

These functional responses are grounded in a complex interplay among charge density waves, topological states~\cite{gooth2019axionic,sehayek2020charge,shi2021charge,chiu2023causal,curtis2023finite,litskevich2024boundary}, lattice chirality~\cite{khandelwal2025topological,kim2025signatures}, superconductivity~\cite{an2020long,mu2021suppression}, structural disorder~\cite{christensen2024disorder}, and structural defects~\cite{huang2021absence}. This multifaceted landscape has motivated various investigations into its temperature-dependent electronic structure, including experimental~\cite{lin2024unconventional} and theoretical~\cite{yang2026microscopic} studies. Although these works have provided valuable insights into the equilibrium properties of the CDW phase, important questions remain about the dynamic processes that govern its phase transitions.

Ultrafast spectroscopic techniques provide a powerful means to investigate the temporal evolution of CDW order, offering femtosecond-scale time resolution. Coherent phonon excitation offers a nonthermal, symmetry-selective pathway to modulate quantum phases in solids~\cite{basov2017towards,nova2019metastable,de2021colloquium}. In particular, mode-selective phonon driving via both infrared~\cite{luo2021light,luo2025symmetry,luo2026extreme} and Raman channels~\cite{vaswani2020light,jiang2023ab} has enabled ultrafast control of symmetry-breaking and topological transitions in a range of materials, including topological insulators~\cite{luo2019ultrafast,yang2020light,luo2023room}, Dirac semimetals~\cite{sun2017coherent,bhandia2024anomalous,cheng2025revealing} and correlated CDW systems~\cite{werdehausen2018coherent,maklar2023coherent,hasaien2025emergent}. These advances underscore the importance of phonon-driven dynamics in future applications in high-speed electronics, spintronics, and quantum information technologies~\cite{liu2014tuning,schoop2016dirac,song2025quantum}. These insights motivate a focused investigation of coherent phonon dynamics in (TaSe$_4$)$_2$I.

Recent ultrafast experiments on this material have revealed a two-step response to optical excitation: rapid polaron melting, followed by modulation of the CDW order and chirality at picosecond timescales~\cite{cheng2024chirality}. While the initial transition is driven by charge excitation and structural reorganization, the subsequent evolution involves mode-selective coherent phonons. However, these coherent phonons are not directly driven through mode-selective nonlinear coupling; rather, they appear to be a natural feedback response arising from the structural reorganization of the CDW state. This raises a compelling question: can symmetry-preserving coherent phonons themselves drive phase transitions in correlated complex materials?

In this work, we perform first-principles simulations to investigate the phonon landscape of (TaSe$_4$)$_2$I in its CDW phase with a commensurate approximate structure. By evaluating the symmetry of each phonon mode and its effect on the electronic structure, we identify nine principal Raman-active modes, each of which preserves the entire crystal symmetry, that strongly suppress the minimum direct gap along the high-symmetry $\Gamma\text{--}\mathrm{Z}$ line to the meV scale, which we use as a convenient indicator of CDW-gap quenching; meanwhile, Brillouin-zone--wide searches reveal the emergence of Weyl nodes at generic \(k\) points, establishing a globally gapless Weyl-semimetallic regime. Among them, we identify the 2.52 THz amplitude mode \(A(18)\) as the most efficient, requiring minimal atomic displacement and most effectively weakening the Ta-chain tetramerization toward a transient restoration of the uniform-chain geometry. By contrast, the Se-dominated Raman modes can also quench the electronic gap and induce a Weyl-semimetallic regime, but they do so with larger required displacements while leaving the underlying CDW lattice distortion largely intact. Furthermore, we explore nonlinear coupling between this Raman mode and low-frequency infrared-active phonons, revealing strong anharmonic modulation of the amplitude-mode potential and assessing the effectiveness of indirect activation. Our results offer a microscopic mechanism for phonon-driven ultrafast control of CDW, and provide predictive guidance for time-resolved pump–probe experiments aiming to dynamically modulate quantum phases in the (TaSe$_4$)$_2$I material.

\section*{Results}
\subsection*{CDW Amplitude Mode}
In the low-temperature CDW phase of (TaSe$_4$)$_2$I, the primitive cell contains 22 atoms (two formula units) and so it has 66 zone-center phonon modes with the following irreducible representations in the D$_2$ point group: $14A \oplus 18B_1 \oplus 18B_2 \oplus 16B_3$. Based on phonon symmetry analysis~\cite{aroyo2006bilbao}, phonon modes with the symmetries of \(B_1\), \(B_2\), and \(B_3\) are both IR-active and Raman-active modes. Those modes with symmetry \(A\) are only Raman-active. Based on our frozen-phonon calculations, we identify nine principal zone-center \(A\) phonon modes that strongly suppress the minimum direct $\Gamma\text{--}\mathrm{Z}$ gap to the meV scale and exhibit Weyl nodes in the Brillouin-zone--wide search. We refer to them as \(A(18)\), \(A(30)\), \(A(33)\), \(A(40)\), \(A(41)\), \(A(44)\), \(A(47)\), \(A(50)\) and \(A(61)\), based on the frequency ordering of the total 66 phonon modes, with their properties summarized in Table~\ref{Table1}. Note that Raman-active \(A\) modes are symmetry-preserving as the associated phonon displacements do not alter the crystal symmetry of space group $F$222. In contrast, distortions induced by modes with \(B_1\), \(B_2\), and \(B_3\) break the symmetry, lowering the space group from $F$222 to $C$2. These symmetry relationships are summarized in Table S1.
\begin{table}[t]
\caption{Selected zone-center Raman-active $A$ modes of (TaSe$_4$)$_2$I in the CDW phase that reach a strongly $\Gamma\text{--}\mathrm{Z}$-gap-suppressed Weyl-semimetallic regime. The nine modes listed here have $A$ symmetry and are Raman-active; their displacements preserve the $F$222 space group. Listed are the mode index, vibrational frequency (Freq), the mode-dependent normal-mode amplitude ($Q_\mathrm{c}$) at which the direct gap along $\Gamma\text{--}\mathrm{Z}$ is reduced to its minimum value $E_G^{\min}$ (a residual gap in the meV range), and the corresponding atomic displacement ($D_\mathrm{c}$). The last two columns give the harmonic coherent-mode energy at $Q_\mathrm{c}$, $E(Q_\mathrm{c})=\frac{1}{2}(2\pi f)^2 Q_\mathrm{c}^2$, evaluated per 22-atom primitive CDW cell, and the corresponding effective temperature rise $\Delta T_\mathrm{eff}=E(Q_\mathrm{c})/(3N k_\mathrm{B})$ ($N=22$) assuming full thermalization of the deposited energy over all lattice degrees of freedom. Except for the \(A(18)\) mode, where the largest displacement occurs on Ta atoms, the other modes exhibit dominant Se atom motion. Additional phonon mode information is provided in Table S1.}
\label{Table1}
\centering
\begin{ruledtabular}
\begin{tabular}{c c c c c c c}
Idx & \makecell{Freq\\(THz)} &
\makecell{$Q_\mathrm{c}$\\(\AA$\sqrt{\mathrm{amu}}$)} &
\makecell{$D_\mathrm{c}$\\(\AA)} &
\makecell{$E_G^{\min}$\\(meV)} &
\makecell{$E(Q_\mathrm{c})$\\(eV)} &
\makecell{$\Delta T_\mathrm{eff}$\\(K)} \\
\hline
18 & 2.52 & 2.5 & 0.098 & 0.6 & 0.08 & 14 \\
30 & 4.07 & 6.5 & 0.279 & 2.3 & 1.44 & 252 \\
33 & 4.23 & 3.6 & 0.152 & 2.0 & 0.47 & 83 \\
40 & 4.82 & 6.6 & 0.287 & 2.2 & 2.08 & 365 \\
41 & 4.84 & 4.2 & 0.176 & 1.4 & 0.85 & 149 \\
44 & 5.08 & 3.8 & 0.162 & 2.6 & 0.76 & 134 \\
47 & 5.34 & 4.8 & 0.180 & 2.3 & 1.35 & 237 \\
50 & 5.51 & 3.3 & 0.134 & 1.3 & 0.68 & 119 \\
61 & 7.84 & 6.5 & 0.261 & 6.5 & 5.33 & 937 \\
\end{tabular}
\end{ruledtabular}
\end{table}

\begin{figure}[t]
    \centering
    \includegraphics[width=3.5in]{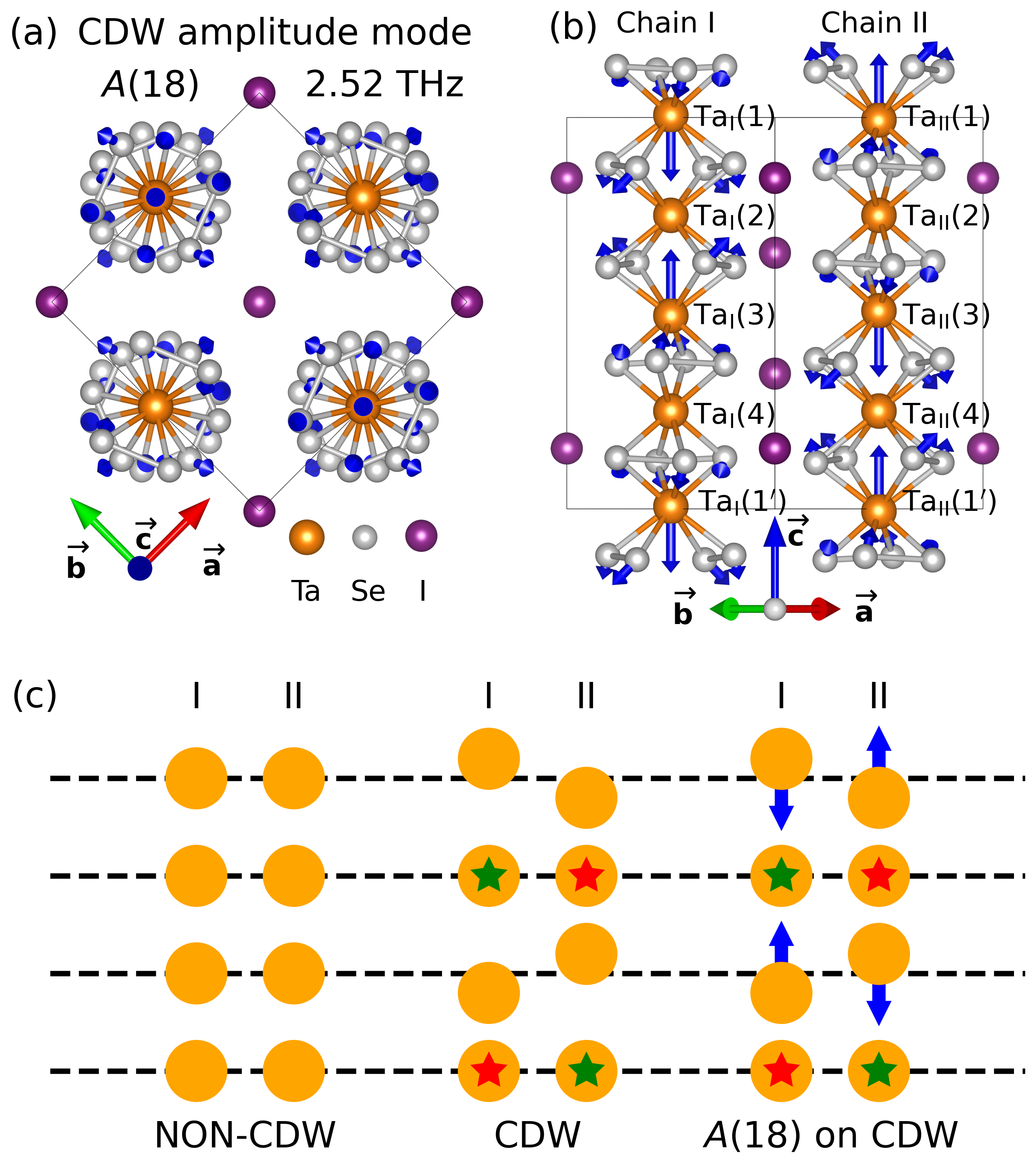}
    \caption{Structure and Ta-chain modulation associated with the CDW amplitude mode $A(18)$.
    (a) Top view and (b) side view of the conventional cell of (TaSe$_4$)$_2$I in the CDW phase. Blue arrows indicate the eigenvectors of the 2.52~THz Raman-active mode \(A(18)\), which preserves the $F222$ symmetry. (c) Schematic of Ta-chain tetramerization in three configurations: the non-CDW state (left), the CDW state (middle), and the CDW state distorted along the \(A(18)\) eigenvector (right). Green stars mark the centers of the \emph{long} Ta--Ta separations on the two chains: Ta$_\mathrm{I}$(1)--Ta$_\mathrm{I}$(3) on chain~I (labeled at Ta$_\mathrm{I}$(2)) and Ta$_\mathrm{II}$(3)--Ta$_\mathrm{II}$(1$^\prime$) on chain~II (labeled at Ta$_\mathrm{II}$(4)). Red stars mark the centers of the \emph{short} Ta--Ta separations: Ta$_\mathrm{II}$(1)--Ta$_\mathrm{II}$(3) on chain~II (labeled at Ta$_\mathrm{II}$(2)) and Ta$_\mathrm{I}$(3)--Ta$_\mathrm{I}$(1$^\prime$) on chain~I (labeled at Ta$_\mathrm{I}$(4)). Here Ta$_\mathrm{I}$(1$^\prime$) and Ta$_\mathrm{II}$(1$^\prime$) denote the periodic images in the neighboring unit cell. Blue arrows illustrate the \(A(18)\)-induced displacement pattern. Orange, gray, and purple spheres denote Ta, Se, and I atoms, respectively.
    }
    \label{Fig1}
\end{figure}

Among these modes, the 2.52 THz Raman mode \(A(18)\), which is the CDW amplitude mode, has the lowest frequency. This mode mainly involves Ta-atoms vibration along the chain direction as shown in Fig.~\ref{Fig1} (a) and (b). The associated Ta atoms movement is the same as Ta-tetramerization of the CDW phase which breaks the symmetry of the isometric chain~\cite{zhang2020first}. As illustrated in Fig.~\ref{Fig1}(c), the non-CDW phase exhibits equal distances between Ta atoms, while the CDW phase shows alternating long and short Ta--Ta distances. The green and red stars in Fig.~\ref{Fig1}(c) mark the centers of the long and short Ta--Ta separations on chains I and II, respectively. Our target amplitude mode \(A(18)\) labeled as blue arrows at right sub-panel of Fig.~\ref{Fig1} (c) modulates the Ta--Ta separations along the chains, reducing the long–short alternation and transiently driving the tetramerized pattern toward the nearly equal-spacing configuration of the non-CDW phase.

\begin{figure*}[t]
    \centering
    \includegraphics[width=6.5in]{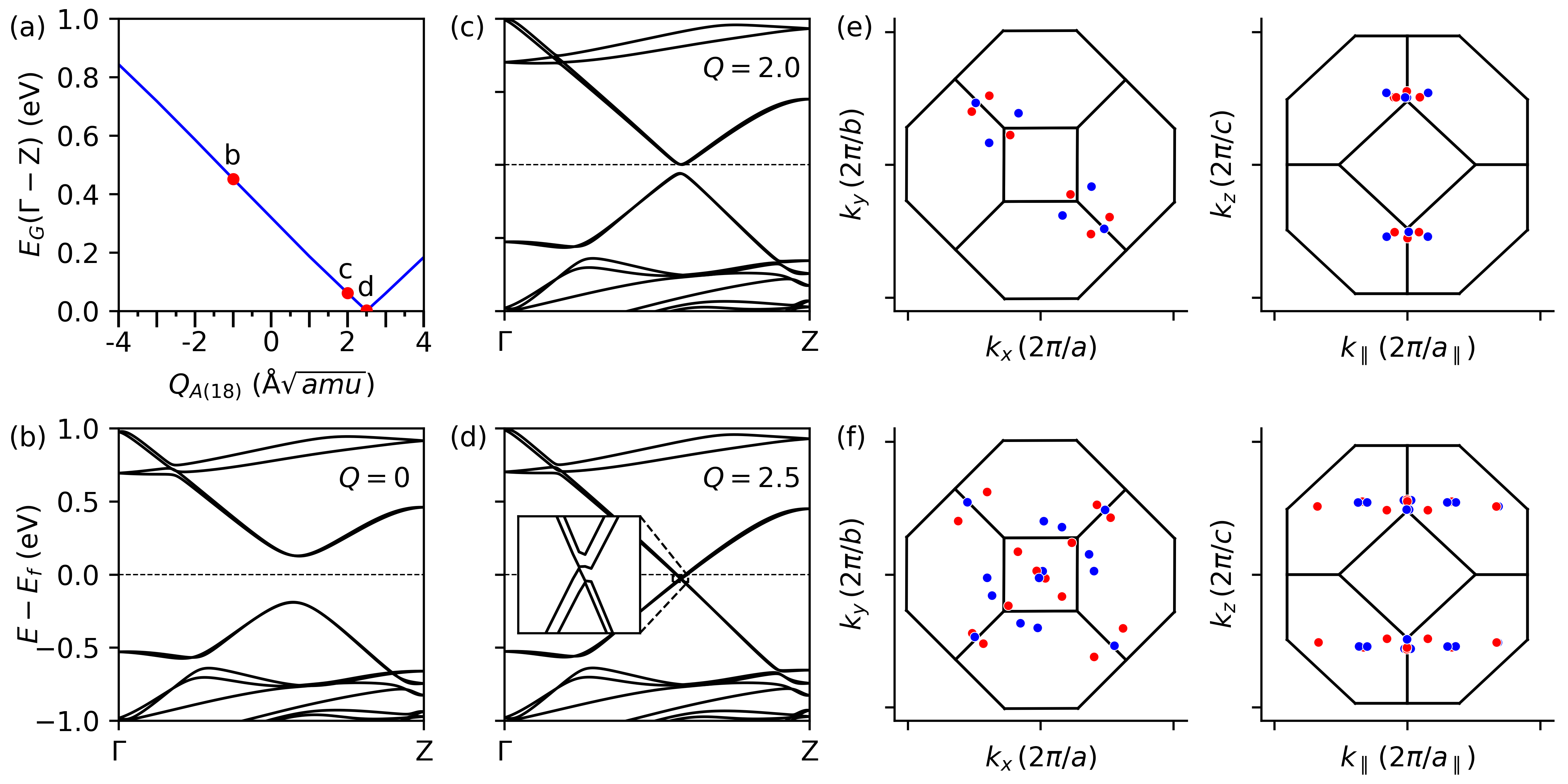}
    \caption{Phonon-driven suppression of the CDW gap and emergence of Weyl nodes under the $A(18)$ mode. (a) Phonon displacement-dependent band gap along $\Gamma\text{--}\mathrm{Z}$ in the CDW phase. Three red circles mark the \(A(18)\) normal-mode amplitudes $Q$ (in $\text{\AA}\sqrt{\mathrm{amu}}$) corresponding to the band structures shown in panels (b) (c) and (d). (b) Electronic band structure along the high-symmetry path $\Gamma\text{--}\mathrm{Z}$ at equilibrium ($Q = 0$), showing a band gap of 0.32~eV. (c) Electronic band structure at $Q = 2.0$, where the Ta--Ta bond disproportionation is nearly removed (the long and short Ta--Ta distances become nearly equal), indicating a transient restoration of the uniform-chain geometry (CDW melting in the structural sense); the $\Gamma\text{--}\mathrm{Z}$ gap is reduced to $60.7$~meV. (d) Electronic band structure at $Q = 2.5$, where the minimum direct gap along $\Gamma\text{--}\mathrm{Z}$ is reduced to a residual direct gap of $0.6$~meV. The inset magnifies the energy window $E-E_F\in[-0.05,-0.01]$~eV to resolve the residual gap. (e) Weyl nodes obtained from a Brillouin-zone--wide search for the phonon-modulated structure at $Q=2.0$, projected onto the $k_x$--$k_y$ plane (left) and the $k_{\parallel}$--$k_z$ plane (right). (f) Same as (e) but for $Q=2.5$. Red and blue spheres denote Weyl nodes with positive and negative chirality, respectively; all nodes shown lie within 10~meV of the Fermi level. Here $k_{\parallel}=(k_x+k_y)/\sqrt{2}$ is the in-plane momentum along the $[110]$ direction, plotted in units of $2\pi/a_{\parallel}$ with $a_{\parallel}=\sqrt{a^2+b^2}$.
    }
    \label{Fig2}
\end{figure*}

The \(A(18)\) mode is experimentally observed at 2.6~THz~\cite{schafer2013dynamics,bronsch2022ultrafast}, consistent with our theoretical prediction. To investigate the electron-phonon coupling effect, we compute the dependence of electronic structure on \(A(18)\) phonon displacement with the frozen-phonon approximation. This treatment should be understood as an adiabatic snapshot approach, in which the electronic structure is evaluated for each instantaneous lattice configuration along the coherent-phonon coordinate \(Q\). Therefore, the \(Q\)-dependent band structure provides an adiabatic description of CDW-gap suppression along the selected coherent-phonon coordinate, whereas the existence of Weyl nodes is determined separately by the full-Brillouin-zone topological analysis. The evolution of the minimum direct gap along the high-symmetry $\Gamma\text{--}\mathrm{Z}$ as a function of the normal mode coordinate $Q$ is shown in Fig.~\ref{Fig2}(a). Here, $Q$ quantifies the atomic displacement along the \(A(18)\) mode, with $Q = 1$~\AA$\sqrt{\text{amu}}$ corresponding to a displacement of 0.039~\AA{} for the dominant Ta$_\mathrm{I,II}$(1,3) atoms. The sign of \(Q\) follows the chosen phase convention of the phonon eigenvector. Here we define positive \(Q\) for \(A(18)\) as the direction that reduces the Ta-chain tetramerization and moves the structure toward the nearly equal-spacing non-CDW geometry, whereas negative \(Q\) corresponds to the opposite phase and enhances the CDW-like long--short Ta--Ta disproportion. With this convention, the asymmetric gap response between \(+Q\) and \(-Q\) reflects the inequivalent electronic effects of weakening and strengthening the CDW distortion. Experimentally, the positive-$Q$ branch is the one expected to be accessed under photoexcitation: in an absorbing CDW system, carrier excitation across the gap reduces the electronic energy gain of the Peierls distortion and displaces the quasi-equilibrium lattice coordinate toward reduced tetramerization, launching a coherent oscillation about this displaced position through the displacive excitation mechanism~\cite{zeiger1992theory}. This mechanism selectively drives totally symmetric ($A$) modes such as $A(18)$, consistent with our focus on the symmetry-preserving Raman modes. We use the minimum direct $\Gamma\text{--}\mathrm{Z}$ gap as the primary operational descriptor of CDW-gap suppression because this path follows the quasi-one-dimensional chain direction in reciprocal space (along the reciprocal-lattice vector $c^{\ast}$) and captures the most pronounced CDW-induced gap opening in the low-temperature phase. Comparisons with other high-symmetry paths show that the $\Gamma\text{--}\mathrm{Z}$ gap is generally among the most sensitive to the lattice distortion amplitude $Q$ and among the fastest to collapse (Figure S1).

At equilibrium ($Q=0$), the band structure exhibits a $\Gamma\text{--}\mathrm{Z}$ gap of 0.32~eV in Fig.~\ref{Fig2}(b). This value is somewhat larger than, but still comparable to, the experimentally reported CDW gap of approximately \(200\sim260\)~meV~\cite{huang2021absence,gooth2019axionic}. Previous first-principles calculations using PBEsol also found that the orthorhombic \(F222\) CDW-I phase opens a gap of approximately \(0.2\)~eV when spin--orbit coupling is included~\cite{zhang2020first}. As the distortion amplitude increases, this $\Gamma\text{--}\mathrm{Z}$ gap gradually decreases and reaches its minimum value of 0.6~meV at $Q = 2.5$~\AA$\sqrt{\text{amu}}$, as shown in Fig.~\ref{Fig2}(d), where the inset resolves the residual opening. Brillouin-zone--wide Weyl-node searches reveal 24 pairs of Weyl nodes at generic \(k\) points in this phonon-modulated structure shown in Fig.~\ref{Fig2}(f). Notably, Fig.~\ref{Fig2}(f) also shows that these nodes are not located exactly on the high-symmetry $\Gamma\text{--}\mathrm{Z}$ line. Upon lowering the symmetry from $I422$ to $F222$, the $C=\pm2$ double Weyl nodes on $\Gamma\text{--}\mathrm{Z}$ are no longer symmetry-protected and are absent in our CDW structure~\cite{shi2021charge,li2021type}. We find only $C=\pm1$ Weyl nodes at generic \(k\) points, indicating that the double-Weyl degeneracy is lifted, potentially accompanied by node splitting and/or annihilation with opposite-chirality partners. 

The strong gap suppression under \(A(18)\) correlates with the modulated Ta–Ta separations. As shown in Table~\ref{Table2}, the \(A(18)\) mode can transiently shorten the Ta$_{\mathrm{I}}$(1)–Ta$_{\mathrm{I}}$(3) separation from 6.635~\AA{} to 6.439~\AA{} at the gap-minimizing displacement, while lengthening the Ta$_{\mathrm{II}}$(1)–Ta$_{\mathrm{II}}$(3) separation from 6.318~\AA{} to 6.514~\AA{}. This brings the pairs of Ta–Ta atoms toward equal distance. However, at $Q = 2.5$~\AA$\sqrt{\text{amu}}$, the Ta–Ta distances are not completely equivalent to those in the non-CDW phase. By monitoring the Ta--Ta separations as a function of $Q$, we find that the chain disproportionation is essentially removed at $Q=2.0~\text{\AA}\sqrt{\text{amu}}$, where the long and short Ta--Ta distances become nearly equal (i.e., a transient restoration of the uniform-chain geometry). At this distortion the $\Gamma\text{--}\mathrm{Z}$ gap has been reduced to 60.7~meV [Fig.~\ref{Fig2}(c)], and a Brillouin-zone--wide Weyl-node search finds 12 pairs of Weyl nodes [Fig.~\ref{Fig2}(e)], indicating that the onset of semimetallic topology does not require the $\Gamma\text{--}\mathrm{Z}$ gap to be fully minimized. Thus, the amplitude for the first Weyl-pair creation is no greater than \(2.0~\mathrm{\AA}\sqrt{\mathrm{amu}}\). Upon increasing the amplitude further to $Q=2.5~\text{\AA}\sqrt{\text{amu}}$, the $\Gamma\text{--}\mathrm{Z}$ gap reaches its minimum and the number of Weyl-node pairs increases to 24 pairs [Fig.~\ref{Fig2}(f)]. 

To estimate the experimental scale of the required coherent-phonon amplitude, we convert the normal-mode coordinate into a harmonic mode energy, \(E_\nu(Q)=\frac{1}{2}(2\pi f_\nu)^2Q^2\). For the \(A(18)\) mode with \(f_\nu=2.52\)~THz, the relevant amplitudes \(Q=2.0\)--\(2.5~\mathrm{\AA}\sqrt{\mathrm{amu}}\) correspond to coherent-mode energies of approximately \(0.05\)--\(0.08\)~eV per primitive CDW cell and maximum Ta displacements of \(0.078\)--\(0.098\)~\AA. If this energy were fully thermalized over the lattice degrees of freedom, it would correspond to an effective temperature increase of order \(10\)--\(15\)~K. Using a representative optical penetration depth of \(0.1\)--\(1~\mu\mathrm{m}\), the corresponding absorbed fluence is estimated to be on the order of \(0.1\)--\(2~\mathrm{mJ/cm^2}\). The actual incident fluence can be higher depending on the absorption coefficient, damping, pulse duration, and coherent-phonon generation efficiency, so these values should be interpreted as order-of-magnitude lower-bound estimates.
The coherent-mode energy $E(Q_\mathrm{c})=\frac{1}{2}(2\pi f)^2 Q_\mathrm{c}^2$ and the corresponding effective temperature rise $\Delta T_\mathrm{eff}=E(Q_\mathrm{c})/(3N k_\mathrm{B})$ ($N=22$), evaluated at each mode's $Q_\mathrm{c}$, are listed in the last two columns of Table~\ref{Table1}. Among the nine Raman modes considered here, $A(18)$ requires the smallest energy input ($\sim0.08$~eV per primitive cell) and the smallest $\Delta T_\mathrm{eff}$ ($\approx14$~K).
To further clarify the microscopic origin, we analyzed element-resolved weights along $\Gamma\text{--}\mathrm{Z}$ and an element-selective decomposition of the $A(18)$ eigenvector (Figures S2 and S3). The near-gap states are dominated by Ta with a secondary Se contribution, while I is negligible. Consistent with this picture, the gap evolution under $A(18)$ is controlled mainly by Ta: when the distortion is restricted to Ta atoms (with the same normalization of $Q$), the minimum direct $\Gamma\text{--}\mathrm{Z}$ gap reaches the meV scale already at $Q=2.0~\text{\AA}\sqrt{\text{amu}}$, compared with $Q=2.5~\text{\AA}\sqrt{\text{amu}}$ for the full mode. By contrast, a Se-only distortion weakly \emph{increases} the $\Gamma\text{--}\mathrm{Z}$ gap for $Q>0$, partially compensating the Ta-driven gap reduction. Accordingly, the $Q_\mathrm{c}$ for the full $A(18)$ mode reflects a balance between Ta- and Se-selective contributions rather than Ta--Ta geometry alone.

\begin{table}[t]
\caption{Interatomic Ta--Ta separations (in \AA) in chain~I and chain~II for the non-CDW, CDW, and \(A(18)\)-modulated structures. The \(A(18)\)-modulated structure corresponds to the frozen-phonon distortion of the 2.52~THz Raman-active mode at \(Q=2.5~\AA\sqrt{\mathrm{amu}}\), where the minimum direct \(\Gamma\text{--}\mathrm{Z}\) gap is reduced to the meV scale.}
\label{Table2}
\centering
\begin{ruledtabular}
\begin{tabular}{lccc}
Ta--Ta separation (\AA) & Non-CDW & CDW & \(A(18)\) ($Q=2.5$)\\
\hline
Ta$_{\mathrm{I}}$(1) -- Ta$_{\mathrm{I}}$(3)   & 6.475 & 6.635 & 6.439 \\
Ta$_{\mathrm{II}}$(1) -- Ta$_{\mathrm{II}}$(3) & 6.475 & 6.318 & 6.514 \\
Ta$_{\mathrm{I}}$(2) -- Ta$_{\mathrm{I}}$(4)   & 6.475 & 6.476 & 6.476 \\
Ta$_{\mathrm{II}}$(2) -- Ta$_{\mathrm{II}}$(4) & 6.475 & 6.476 & 6.476 \\
\end{tabular}
\end{ruledtabular}
\end{table}

\subsection*{Se-Dominated Raman Modes}
\begin{figure*}[t]
    \centering
    \includegraphics[width=7.0in]{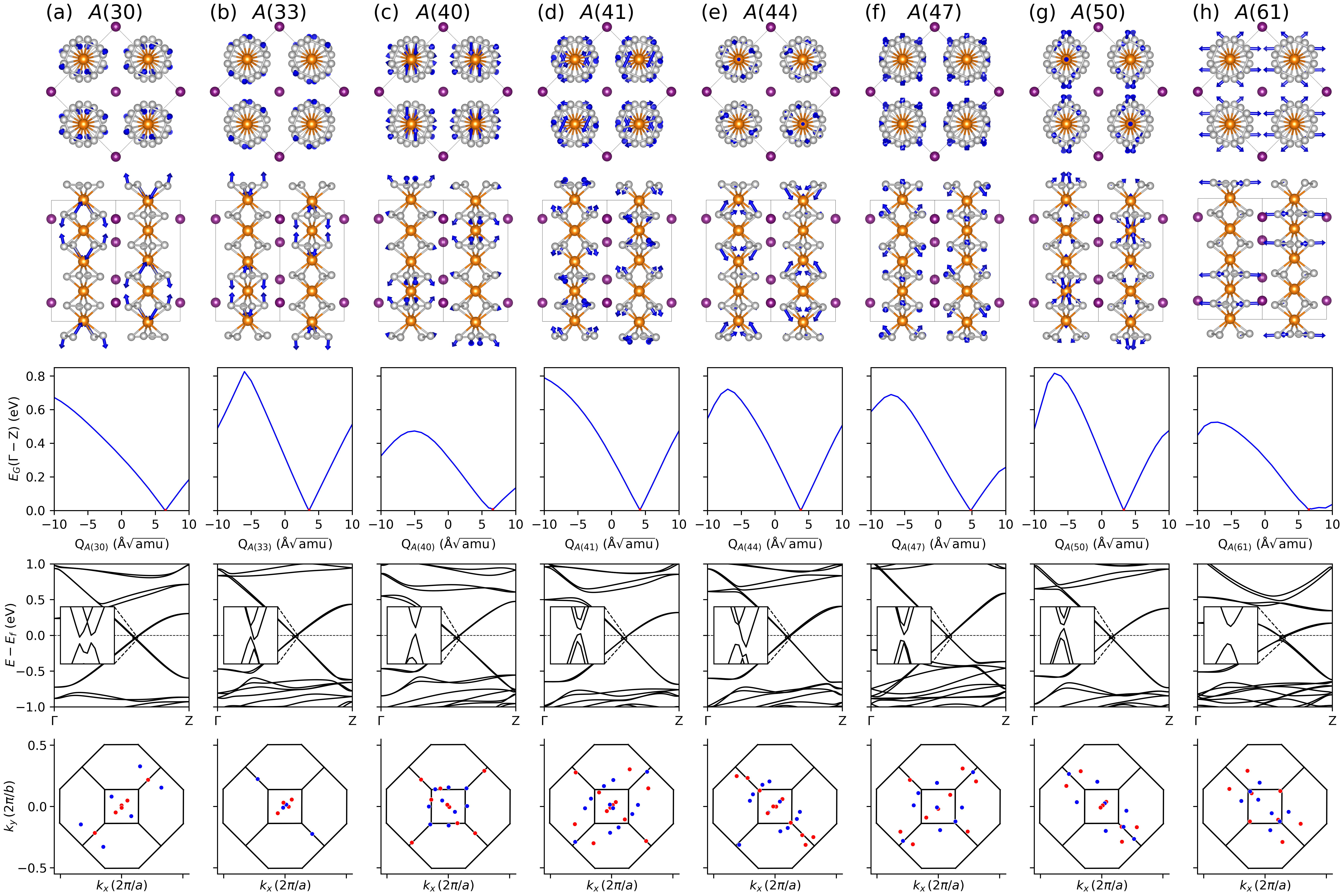}
    \caption{Phonon-driven $\Gamma\text{--}\mathrm{Z}$ gap suppression in (TaSe$_4$)$_2$I by Raman-active modes beyond the CDW amplitude mode. (a)--(h) Results for eight additional principal Raman-active modes: $A(30)$, $A(33)$, $A(40)$, $A(41)$, $A(44)$, $A(47)$, $A(50)$, and $A(61)$, respectively. For each mode, the \emph{first} and \emph{second} panels show the phonon eigenvectors in the top and side views, respectively; blue arrows indicate the atomic displacement directions, with dominant motion observed on Se atoms. The \emph{third} panels show the evolution of the band gap along the $\Gamma\text{--}\mathrm{Z}$ path as a function of the normal-mode coordinate $Q$, showing that each mode can suppress the gap to the meV scale at a mode-dependent near-closure amplitude $Q_\mathrm{c}$ (Table~\ref{Table1}). The \emph{fourth} panels present corresponding electronic band structures along $\Gamma\text{--}\mathrm{Z}$ at $Q=Q_\mathrm{c}$, showing a near-touching along $\Gamma\text{--}\mathrm{Z}$ with a meV-scale residual gap. Insets zoom into the near-$E_F$ energy range to resolve the meV-scale residual opening; the energy window is chosen individually for each mode for clarity. Here, $E_F$ denotes the Fermi level. The \emph{fifth} panels show Brillouin-zone--wide Weyl-node searches for the phonon-distorted structures at $Q=Q_\mathrm{c}$, showing that the Weyl nodes occur at generic $k$ points rather than exactly on the $\Gamma\text{--}\mathrm{Z}$ line. Red and blue spheres denote Weyl nodes with positive and negative chirality, respectively.}
    \label{Fig3}
\end{figure*}

In addition to the CDW amplitude mode \(A(18)\), we focus on eight other principal Raman-active phonon modes that can strongly suppress the minimum direct gap along the high-symmetry $\Gamma\text{--}\mathrm{Z}$ line and drive the system into a globally gapless Weyl-semimetallic regime. In Fig.~\ref{Fig3}, for each mode, the first and second panels show the eigenvectors from the top and side views, respectively, and the third panel shows the evolution of the minimum direct $\Gamma\text{--}\mathrm{Z}$ gap as a function of the normal-mode coordinate $Q$. The corresponding electronic structures at the gap-minimizing (near-gap-closure) displacements are shown in the fourth panel. Each mode exhibits a distinct amplitude $Q_\mathrm{c}$ required to reach the near-closure condition, as summarized in Table~\ref{Table1}. Along the high-symmetry $\Gamma\text{--}\mathrm{Z}$ line, the near-gap bands approach each other closely, leaving only a meV-scale residual gap, indicating that the near-closure occurs in the vicinity of this line. Meanwhile, Brillouin-zone--wide node searches identify Weyl nodes at generic \(k\) points (fifth panel) rather than exactly on $\Gamma\text{--}\mathrm{Z}$. This is expected because the \(F222\) CDW structure already lacks the \(C_{4z}\) symmetry of the parent phase that protects the \(C=\pm2\) double-Weyl degeneracies on the \(\Gamma\text{--}\mathrm{Z}\) line, while the symmetry-preserving \(A\)-mode distortions retain the \(F222\) space group. Consistent with this symmetry setting, the Weyl nodes obtained in our calculations occur at generic \(k\) points. In particular, mode~A(50) shows an especially strong near-closure along $\Gamma\text{--}\mathrm{Z}$, with a residual gap of 1.3~meV, and the resulting Weyl nodes are found very close to this line, although they remain slightly off $\Gamma\text{--}\mathrm{Z}$ when their coordinates are examined.

Notably, unlike the CDW amplitude mode \(A(18)\), these Se-dominated distortions do not necessarily produce a pronounced reduction of the Ta-chain tetramerization, suggesting that electronic gap quenching can occur even when the underlying CDW lattice distortion largely persists. A common characteristic across all eight Se-dominated modes is that the largest atomic displacements primarily originate from Se vibrations. However, the required maximum atomic displacements $D_\mathrm{c}$ are substantially larger than that of the CDW amplitude mode, and these modes span higher frequencies from \(4.07\) to \(7.84\)~THz. In fact, Table~\ref{Table1} shows that several Se-dominated modes (e.g., \(A(30)\), \(A(40)\), and \(A(61)\)) require $D_\mathrm{c}\gtrsim 0.2$~\AA{}, i.e., lattice-scale displacements that are likely difficult to achieve coherently without substantial heating or structural damage. Across all eight Se-dominated modes, the harmonic coherent-mode energies range from \(0.47\) to \(5.33\)~eV per primitive CDW cell, corresponding to energy-equivalent temperature scales of \(83\)--\(937\)~K, compared with \(0.08\)~eV and \(14\)~K for \(A(18)\). We emphasize that \(\Delta T_{\mathrm{eff}}\) is an energy-equivalent scale rather than a prediction of the actual pump-induced sample temperature. This comparison supports the conclusion that \(A(18)\) is the most energetically favorable Raman pathway among the modes considered.
Therefore, in the next section discussing IR-coupled Raman processes, we primarily focus on the CDW amplitude mode, as it presents a more experimentally feasible path for driving the insulator-to-semimetal transition~\cite{kim2021terahertz}.

\subsection*{Nonlinear Phonon Coupling Between CDW Amplitude Mode and IR Modes}
Nonlinear phononic coupling has emerged as a powerful mechanism for manipulating lattice dynamics and electronic phases in quantum materials~\cite{forst2011nonlinear}. Recent studies have shown that intense THz excitation can trigger coherent energy transfer across vibrational modes due to mode-selective coupling, enabling ultrafast control of structural and electronic properties~\cite{cheng2019helicity,liu2020coherent,liu2020ultrafast,liu2024coherent}. A particularly important class of such interactions involves the excitation of an IR-active phonon that drives a unidirectional displacement in a Raman-active mode via anharmonic coupling. This IR–Raman mechanism forms the basis for dynamical control of structural and electronic phase transitions, including magnetic and topological states, as demonstrated in recent theoretical studies~\cite{khalsa2018ultrafast,feng2022anti,tang2023light}.

In this section, we investigate the nonlinear phononic mechanism in (TaSe$_4$)$_2$I by identifying IR-active modes that couple anharmonically to the Raman-active CDW amplitude mode \(A(18)\). This low-frequency Raman mode is primarily responsible for strongly suppressing the CDW gap and restoring Weyl semimetal features, making it a central target for coherent phonon control. To quantify the coupling, we performed total energy calculations as a function of an IR mode and \(A(18)\) normal coordinates to construct two-dimensional energy surfaces. These potential energy surfaces were then fitted using a fourth-order anharmonic expansion that explicitly captures mode–mode interactions and nonlinear effects~\cite{subedi2014theory,subedi2015proposal,fechner2016effects}. The normal-mode and energy-normalization conventions used in the fitting are detailed in the Supplementary Information~\cite{SM}. The potential is written as:
\begin{align}
V(Q_{\text{IR}}, Q_{\text{R}}) = \frac{K_{\text{IR}}}{2} Q_{\text{IR}}^2 + \frac{K_{\text{R}}}{2} Q_{\text{R}}^2 &&  \notag\\
+ \frac{c_1}{3} Q_{\text{IR}}^3 + \frac{c_2}{3} Q_{\text{R}}^3 + c_{12} Q_{\text{IR}}Q_{\text{R}}^2 + c_{21} Q_{\text{IR}}^2 Q_{\text{R}} && \notag\\ + \frac{d_1}{4} Q_{\text{IR}}^4 + \frac{d_2}{4} Q_{\text{R}}^4 + d_{13} Q_{\text{IR}}Q_{\text{R}}^3 + d_{31} Q_{\text{IR}}^3 Q_{\text{R}} + d_{22} Q_{\text{IR}}^2 Q_{\text{R}}^2 
\end{align}
where $Q_{\text{IR}}$ is the amplitude of an IR mode, and $Q_{\text{R}}$ is the amplitude of the Raman mode. The first two terms represent the harmonic potentials for the IR and Raman normal modes, with $K_{\text{IR}}$ and $K_{\text{R}}$ being the corresponding harmonic coefficients expressed using the same per-atom energy normalization and normal-coordinate convention as the fitted anharmonic coefficients. The cubic terms ($c_1$, $c_2$) describe intrinsic intra-mode anharmonicity, while $c_{12}$ and $c_{21}$ represent third-order IR–Raman coupling strengths. The quartic terms ($d_1$, $d_2$) account for higher-order intra-mode coupling, and $d_{13}$, $d_{31}$, and $d_{22}$ describe fourth-order anharmonic coupling between the modes. 
By symmetry, the lattice potential must be invariant under the operations of the reference structure. Because the IR mode considered here has \(B\) symmetry, its normal coordinate changes sign under the symmetry operation(s) that leave the \(A\)-symmetry Raman coordinate unchanged. Therefore, all terms containing an odd power of $Q_{\mathrm{IR}}$ are forbidden, i.e., 
$c_1=c_{12}=d_{13}=d_{31}=0$, consistent with our numerical fits. The potential thus simplifies to
%Since the potential energy is a scalar, the coupling between the IR B-mode and Raman A-mode is constrained to have vanishing coefficients for terms with odd power of \(Q_\textrm{IR}\). Therefore we have \(c_1=c_{12}=d_{13}=d_{31}=0\), as also confirmed numerically. So the potential is simplied to 
\begin{align}
\label{eq:V_IR_R}
V(Q_{\text{IR}}, Q_{\text{R}}) = \frac{K_{\text{IR}}}{2} Q_{\text{IR}}^2 + \frac{K_{\text{R}}}{2} Q_{\text{R}}^2 &&  \notag\\
+ \frac{c_2}{3} Q_{\text{R}}^3 + c_{21} Q_{\text{IR}}^2 Q_{\text{R}} + \frac{d_1}{4} Q_{\text{IR}}^4 + \frac{d_2}{4} Q_{\text{R}}^4 + d_{22} Q_{\text{IR}}^2 Q_{\text{R}}^2 
\end{align}

In particular, the $c_{21} Q_{\mathrm{IR}}^2 Q_{\mathrm{R}}$ and $d_{22} Q_{\mathrm{IR}}^2 Q_{\mathrm{R}}^2$ terms describe, respectively, the rectified force on the Raman coordinate and the modulation of the Raman potential curvature. The $c_{21} Q_{\mathrm{IR}}^2 Q_{\mathrm{R}}$ term enables rectification of the oscillatory IR mode into a directional force on the Raman mode, as discussed in theoretical proposals for light-induced structural transitions~\cite{subedi2014theory,subedi2015proposal}. Meanwhile, the $d_{22} Q_{\mathrm{IR}}^2 Q_{\mathrm{R}}^2$ term modulates the phonon energy landscape and can enhance or suppress the IR--Raman interaction depending on the mode symmetry and excitation strength, as demonstrated in studies of magnetophononic switching and light-induced phase control~\cite{fechner2016effects}.

We computed the anharmonic coupling coefficients between all IR-active modes and the Raman-active CDW amplitude mode \(A(18)\). The complete dataset is presented in Table S2, while the results of several representative IR modes are summarized in Table~\ref{Table3}. For example, \(B_2(8)\), \(B_1(9)\) (both near 1.3~THz), and \(B_3(17)\)~\cite{cheng2024chirality} exhibit weak anharmonic coupling with \(A(18)\), as evidenced by their small $c_{21}$ and $d_{22}$ coefficients. In contrast, \(B_3(7)\), \(B_3(34)\), \(B_3(42)\), and \(B_1(54)\) show significantly stronger anharmonic interactions, particularly through the $c_{21}$ and $d_{22}$ terms. Among these, \(B_3(7)\) stands out due to its especially large $d_{22}$ coefficient. Moreover, its low vibrational frequency makes \(B_3(7)\) energetically favorable for excitation, identifying it as a promising candidate for nonlinear modulation of the CDW amplitude mode potential through IR--Raman anharmonic coupling.

\begin{table}[t]
\caption{Anharmonic coupling coefficients between selected IR-active phonon modes and the Raman-active CDW amplitude mode \(A(18)\) in (TaSe$_4$)$_2$I. Listed are the mode indices, frequencies (THz), irreducible representations, and symmetry-allowed third- and fourth-order coefficients (\(c_2\), \(c_{21}\), \(d_1\), \(d_2\), \(d_{22}\)) extracted from the nonlinear energy expansion in Eq.~(\ref{eq:V_IR_R}). Energies used in the fitting are in meV per atom; coefficients are given in \(\mathrm{meV}\,\mathrm{atom}^{-1}/\bigl(\mathrm{amu}^{k/2}\,\text{\AA}^{k}\bigr)\), where \(k\) is the total order of displacements (\(k=3\) for cubic and \(k=4\) for quartic terms).}
\label{Table3}
\centering
\begin{ruledtabular}
\begin{tabular}{r r c r r r r r}
Idx & Freq & Irrep & \(c_2\) & \(c_{21}\) & \(d_1\) & \(d_2\) & \(d_{22}\) \\
\hline
7  & 1.15 & \(B_3\) & -0.180 & -0.080 & -0.002 & -0.019 &  0.009 \\
8  & 1.27 & \(B_2\) & -0.190 &  0.005 &  0.020 & -0.019 & -0.002 \\
9  & 1.30 & \(B_1\) & -0.190 &  0.003 &  0.034 & -0.019 & -0.002 \\
17 & 2.01 & \(B_3\) & -0.189 & -0.001 &  0.032 & -0.019 & -0.001 \\
34 & 4.30 & \(B_3\) & -0.189 &  0.077 &  0.036 & -0.019 &  0.000 \\
42 & 4.91 & \(B_3\) & -0.191 & -0.085 &  0.040 & -0.018 & -0.005 \\
54 & 5.87 & \(B_1\) & -0.190 & -0.068 &  0.034 & -0.019 & -0.005 \\
\end{tabular}
\end{ruledtabular}
\end{table}

\begin{figure}[t]
    \centering
    \includegraphics[width=3.5in]{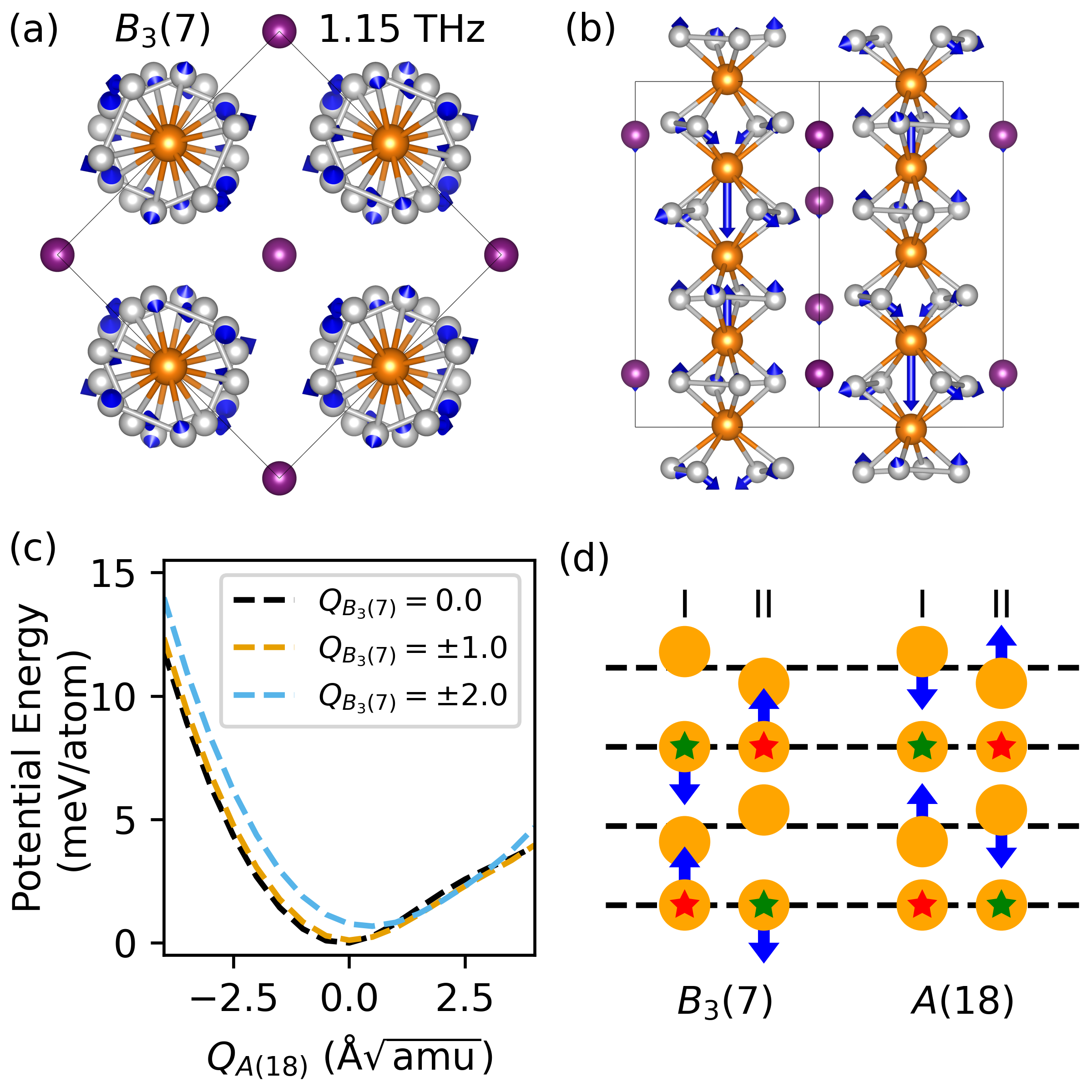}
    \caption{
    Nonlinear coupling between the IR-active mode \(B_3(7)\) and the CDW amplitude mode \(A(18)\). (a) and (b) show the top and side views of the atomic displacements (eigenvectors) for the \(B_3(7)\) mode, respectively, highlighting Ta atom vibrations along the c-axis, similar to the \(A(18)\) mode. (c) Anharmonic potential energy surface of the Raman \(A(18)\) mode modulated by the \(B_3(7)\) IR mode. The three curves correspond to three different amplitudes of the \(B_3(7)\) mode: the black dashed line represents $Q_{B_3(7)} = 0.0$, showing the baseline potential energy curve of \(A(18)\) in the absence of IR distortion; the orange dashed line corresponds to $Q_{B_3(7)} = \pm 1.0$, where the energy landscape slightly stiffens; and the blue dashed line corresponds to $Q_{B_3(7)} = \pm 2.0$, showing a noticeable increase in curvature and a slight shift of the potential minimum toward positive \(Q_{A(18)}\). (d) Schematic comparison between the \(B_3(7)\) and \(A(18)\) modes, demonstrating the additional two-fold symmetry breaking caused by \(B_3(7)\) mode displacement. Red and green stars mark the centers of the short and long Ta--Ta separations, respectively, and blue arrows indicate the phonon-induced Ta displacements.
    }
    \label{Fig4}
\end{figure}

The eigenvectors of the \(B_3(7)\) mode are shown in Fig.~\ref{Fig4}(a) and (b) from the top and side views, respectively. This mode primarily involves vibrations of Ta atoms along the c-axis, similar to the Raman \(A(18)\) mode. A direct comparison between \(B_3(7)\) and \(A(18)\) is illustrated in Fig.~\ref{Fig4}(d). Note that \(B_3(7)\) breaks the two-fold rotation symmetries about the \(x\) and \(y\) axes and lowers the space group from \(F222\) to \(C2\), whereas the \(A(18)\) distortion preserves these symmetries and leaves the space group unchanged ($F222$)(see Table S1).

We can further demonstrate the anharmonic coupling between the \(A(18)\) Raman mode and \(B_3(7)\) IR mode by plotting the potential energy curve of \(A(18)\) at finite \(B_3(7)\) displacements, as presented in Fig.~\ref{Fig4}(c). The black dashed line with the normal mode coordinate $Q_{B_3(7)} = 0$ represents the baseline potential energy curve of \(A(18)\) in the absence of IR distortion. The orange dashed line ($Q_{B_3(7)} = \pm1.0$) shows a slight stiffening of the energy profile, while the blue dashed line ($Q_{B_3(7)} = \pm2.0$) reveals a noticeably steeper curvature.

These results show that excitation of the \(B_3(7)\) IR-active mode significantly modifies the potential energy profile of the \(A(18)\) Raman mode, confirming the coupling is comparatively strong among the IR--\(A(18)\) couplings considered. As the \(B_3(7)\) amplitude increases, the \(A(18)\) potential energy curve becomes progressively steeper, consistent with the comparatively large fourth-order coupling coefficient $d_{22}$. This establishes \(B_3(7)\) as a promising candidate for nonlinear modulation of the CDW amplitude mode through selective IR excitation and Raman mode-coupling in pump–probe experiments.

More specifically, we observe a slight shift of the \(A(18)\) potential minimum toward positive $Q_{\text{R}}$ values as the \(B_3(7)\) amplitude increases. This asymmetric shift is consistent with the third-order coupling term $c_{21} Q_{\text{IR}}^2 Q_{\text{R}}$ with \(c_{21} =-0.080\), which rectifies the IR phonon oscillation into a net directional force on the Raman mode~\cite{subedi2014theory,subedi2015proposal}. In contrast, the quartic $d_{22} Q_{\text{IR}}^2 Q_{\text{R}}^2$ term governs the curvature modulation, leading to the observed stiffening of the energy profile, as illustrated in Fig.~\ref{Fig4}(c). For a coherently oscillating IR mode, the leading cycle-averaged rectified force is \(\overline{F}_{\mathrm{R}}(0) =-c_{21}\langle Q_{\mathrm{IR}}^2\rangle\). Because \(c_{21}<0\), the force points toward positive \(Q_{\mathrm{R}}\), i.e., toward the CDW-suppressing structural direction. The detailed derivation and quantitative quasi-static estimate are provided in the Supplementary Information~\cite{SM}.

These qualitative insights are quantitatively supported by the coupling coefficients listed in Table~\ref{Table3}. Among all examined IR modes, \(B_3(7)\) exhibits the largest positive $d_{22}$ coefficient and a sizable negative $c_{21}$, consistent with its influence on both curvature stiffening and the shift of the potential minimum toward positive \(Q_{\mathrm{R}}\). The consistency between the energy landscape in Fig.~\ref{Fig4}(c) and the extracted coefficients provides a coherent picture of anharmonic coupling and highlights \(B_3(7)\) as the most favorable low-frequency IR-active mode considered here for nonlinear modulation of the CDW amplitude mode potential.

To assess experimental feasibility, we estimate that, under favorable resonant conditions and using an illustrative phonon quality-factor range \(\mathcal{Q}_{\mathrm{ph}}=10\)--\(40\), a representative peak amplitude \(Q_0=2.0~\mathrm{\AA}\sqrt{\mathrm{amu}}\) of the driven \(B_3(7)\) coordinate corresponds to a steady-state THz field scale of approximately \(15\)--\(60~\mathrm{kV/cm}\). Finite-duration pulses may require larger peak fields; details of the mode-effective-charge calculation, field estimate, and quasi-static rectification analysis are provided in the Supplementary Information~\cite{SM}.

However, at this benchmark amplitude, the complete quasi-static estimate gives a cycle-averaged displacement of \(\overline{Q_R^{\mathrm{ind}}}\simeq+0.134~\mathrm{\AA}\sqrt{\mathrm{amu}}\). Its magnitude is only approximately \(5.4\%\)--\(6.7\%\) of the positive-branch \(A(18)\) amplitudes \(Q_R=2.0\)--\(2.5~\mathrm{\AA}\sqrt{\mathrm{amu}}\) relevant for strong suppression of the Ta-chain bond disproportionation and the CDW gap, corresponding to a magnitude smaller by a factor of approximately fifteen to nineteen. Importantly, the induced displacement has the favorable sign and biases \(A(18)\) toward the CDW-suppressing structural branch. The quasi-static estimate describes only the cycle-averaged static component of the response; the oscillatory component at twice the driven IR-mode frequency may produce an additional transient \(A(18)\) response that is not included here. Thus, although \(B_3(7)\) assists or biases the \(A(18)\) coordinate in the favorable direction, the resulting cycle-averaged displacement remains far too small for \(B_3(7)\) excitation alone to provide a standalone pathway to the Weyl-semimetallic regime.

\section*{Discussion}
In this work, we have presented a detailed first-principles study of coherent phonon-driven control of the CDW state in the quasi-one-dimensional material (TaSe$_4$)$_2$I. By systematically exploring the zone-center phonon modes of its low-temperature CDW phase, we identified nine principal symmetry-preserving Raman-active modes that strongly suppress the CDW gap and access a topologically nontrivial Weyl-semimetallic regime. Among these, the 2.52~THz amplitude mode \(A(18)\) emerges as the most efficient driver of this transition, requiring minimal atomic displacement to suppress the $\Gamma\text{--}\mathrm{Z}$ gap to the meV scale. This mode primarily modulates the Ta-tetramer chain structure, driving it toward a geometry closely resembling that of the non-CDW phase and enabling the reappearance of Weyl points.

While several higher-frequency Raman modes dominated by Se vibrations can likewise suppress the $\Gamma\text{--}\mathrm{Z}$ gap to the meV scale, they typically do so without substantially reducing the Ta-chain tetramerization, so that the underlying CDW lattice distortion largely remains intact. Moreover, these modes generally require significantly larger atomic displacements and are therefore less accessible under experimental conditions. This highlights the special role of \(A(18)\) as an experimentally feasible and symmetry-compatible driver of topological switching.

From a topological perspective, the phonon-driven suppression of the $\Gamma\text{--}\mathrm{Z}$ gap alone does not establish a global topological transition. We therefore use it only as an operational descriptor of CDW-gap suppression, whereas the existence of the globally gapless Weyl-semimetallic regime is established independently by Brillouin-zone--wide Weyl-node searches. For the 2.52~THz Raman mode \(A(18)\), the BZ-wide search reveals Weyl nodes at generic \(k\) points; at the gap-minimizing distortion we identify 24 pairs of Weyl nodes. By contrast, the Se-dominated Raman $A$ modes can also induce a Weyl-semimetallic regime, but typically with fewer Weyl-node pairs than \(A(18)\), underscoring the exceptional efficiency of the CDW amplitude mode for lattice-driven topological restoration even when the $\Gamma\text{--}\mathrm{Z}$ gap remains finite at the meV scale. We note that resolving the precise boundaries of the Weyl-semimetallic region---the amplitude at which the first Weyl pair is created and that at which the nodes annihilate---would require repeating the full Wannierization and Brillouin-zone--wide node search on a dense grid of amplitudes for each mode, which is computationally prohibitive and left for future work.

Beyond identifying direct Raman pathways for the CDW-insulator-to-Weyl-semimetal transition, we investigated nonlinear phonon–phonon interactions and uncovered strong anharmonic coupling between \(A(18)\) and the low-frequency IR-active mode \(B_3(7)\) at 1.15~THz. This IR–Raman interaction includes a third-order term $c_{21} Q_{\mathrm{IR}}^2 Q_{\mathrm{R}}$, which produces a rectified shift of the Raman-mode equilibrium coordinate, and a fourth-order term $d_{22} Q_{\mathrm{IR}}^2 Q_{\mathrm{R}}^2$, which modifies the curvature of the \(A(18)\) potential energy surface. Our analysis further indicates that the \(B_3(7)\) coupling biases \(A(18)\) toward the CDW-suppressing structural direction, but the estimated indirect response remains far below that required for strong gap suppression. Thus, \(B_3(7)\) may assist the \(A(18)\) response, while direct coherent excitation of \(A(18)\) remains the most effective pathway to the Weyl-semimetallic regime. Our predictions can be experimentally tested using state-of-the-art ultrafast nonlinear spectroscopies, especially recently developed THz pump–probe~\cite{yang2019lightwave,mootz2025observation} and THz two-dimensional coherent spectroscopy techniques~\cite{mootz2022visualization,luo2023quantum,huang2026terahertz2d}. In particular, the transient lattice distortion can be benchmarked against calculated frozen-phonon diffraction patterns, following the approach recently proposed for displacement-driven CDW order~\cite{zhang2024understanding}, while the Weyl-semimetallic regime associated with direct \(A(18)\)
excitation can be independently probed by time-resolved ARPES, the longitudinal circular photogalvanic effect, or anomalous THz-frequency Hall conductivity. These approaches have proven highly effective for probing collective modes~\cite{cheng2024evidence,huang2024extreme,huang2025discovery} and coherent excitations~\cite{yang2018terahertz} across a wide range of quantum materials. 

Our study demonstrates a phonon-mode-selective route for nonthermal modulation of electronic topology in a CDW system. It provides quantitative metrics—including phonon frequencies, displacement amplitudes, and anharmonic coupling coefficients—for guiding experimental access to ultrafast phase transitions. Furthermore, the IR–Raman analysis presented here delineates both the opportunities and limitations of hybrid phonon-driving schemes and provides quantitative guidance for identifying coupling channels with favorable rectification direction and sufficient attainable amplitude.

Together, these findings bridge lattice dynamics, topological band theory, and ultrafast condensed matter physics. They offer predictive guidance for coherent phase engineering in quantum materials, emphasizing the utility of low-frequency phonons and nonlinear coupling mechanisms. While our analysis focuses on (TaSe$_4$)$_2$I, the methodology and physical insights are broadly applicable to low-symmetry or low-dimensional quantum materials with strong electron–phonon coupling. Promising directions for future work include real-time phonon–electron simulations, ultrafast spectroscopy to validate IR–Raman activation pathways, and extension of this approach to symmetry-breaking phases such as magnetism, ferroelectricity, and superconductivity. These developments open new opportunities for designing THz-driven topological quantum devices.

\section*{Methods}

\subsection*{DFT and Structural Models}

First-principles density functional theory (DFT) calculations were performed using the projector augmented-wave (PAW) method~\cite{PAW} and the Perdew--Burke--Ernzerhof (PBE) exchange-correlation functional~\cite{PBE}, as implemented in the VASP package~\cite{VASP}. (TaSe$_4$)$_2$I crystallizes in a tetragonal chiral structure with space group $I$422 (No. 97) under ambient conditions, with experimental lattice parameters $a = 9.531$~\AA{} and $c = 12.824$~\AA{}, corresponding to the non-CDW phase~\cite{TaSe8parameterAandC}. Structural optimization was carried out using a $\Gamma$-centered $3 \times 3 \times 2$ $k$-point mesh and a plane-wave cutoff energy of 400 eV, relaxing both lattice constants and atomic positions. The convergence criteria were set to $10^{-8}$ eV for electronic self-consistency and 0.001 eV/\AA{} for ionic relaxation. The optimized lattice parameters for the tetragonal phase are $a = 9.769$~\AA{} and $c = 12.950$~\AA{}. We approximate the incommensurate CDW phase using an orthorhombic structure in space group $F222$ (No. 22), with optimized lattice parameters $a \approx b \approx 9.779$~\AA{} and $c = 12.953$~\AA{}, indicating a nearly tetragonal metric. Selected phonon and frozen-phonon calculations were additionally repeated using the PBEsol exchange-correlation functional~\cite{perdew2008restoring} to assess the robustness of the results against the choice of functional.

\subsection*{Phonon and Symmetry Analysis}

A conventional cell containing 44 atoms (four formula units) was used to compute $\Gamma$-point phonons using the finite-displacement method implemented in PHONOPY~\cite{phonopy}. Frozen-phonon band-structure calculations were performed in the primitive cell by displacing atoms along selected $\Gamma$-point phonon eigenvectors and computing the corresponding band structure using a $\Gamma$-centered $6\times6\times6$ $k$-point mesh. To assess the dynamical stability of the relaxed $F222$ CDW structure, the full phonon dispersion was calculated using a $2\times2\times2$ supercell of the 22-atom primitive cell, corresponding to 176 atoms, with forces evaluated using a $\Gamma$-centered $2\times2\times2$ $k$-point mesh.

For the nonlinear phonon-coupling analysis, two-dimensional frozen-phonon energy surfaces were generated by simultaneously displacing the structure along selected infrared-active and Raman-active phonon eigenvectors and fitting the resulting DFT energies to the symmetry-allowed fourth-order potential used in the main text. The normal coordinates follow the eigenvector normalization of the 22-atom primitive CDW cell, while the calculated energy differences were normalized per atom. Born effective-charge tensors were calculated using density-functional perturbation theory for the relaxed $F222$ CDW structure, and mode effective charges were obtained by projecting the atom-resolved Born effective-charge tensors onto the corresponding normalized phonon eigenvectors.

The crystal symmetry analysis was carried out using the spglib library~\cite{togo2024spglib} through its interface in pymatgen~\cite{ong2013python}. Optical activity analysis was carried out using the Bilbao Crystallographic Server~\cite{aroyo2006bilbao}. Spin--orbit coupling was included in all calculations.

\subsection*{Wannier and Weyl-Node Calculations}

Weyl points were computed by constructing a tight-binding model using maximally localized Wannier functions~\cite{marzari1997maximally,souza2001maximally}. The spinor Wannier Hamiltonian was constructed using Ta $d$, Se $p$, and I $p$ orbital projections, yielding 148 Wannier functions. Weyl nodes were located by an explicit search over the full three-dimensional Brillouin zone, and the chiral charge of each node was obtained from the Berry flux on an enclosing surface, as implemented in the WannierTools package~\cite{wu2018wanniertools}.

\section*{Data Availability}
Data supporting the findings of this study are available at
\href{https://gitlab.com/tjianglab/ta2se8i_raman_driven_weyl}{https://gitlab.com/tjianglab/ta2se8i\_raman\_driven\_weyl}.

\section*{Code Availability}
Custom scripts used to generate and analyze the data and figures in this study are available at
\href{https://gitlab.com/tjianglab/ta2se8i_raman_driven_weyl}{https://gitlab.com/tjianglab/ta2se8i\_raman\_driven\_weyl}.
Software packages and calculation parameters used in this study are described in the Methods and Supplementary Information.

\section*{Acknowledgements}
This work was supported by the U.S. Department of Energy (DOE), Office of Science, Basic Energy Sciences, Materials Science and Engineering Division, including the grant of computer time at the National Energy Research Scientific Computing Center (NERSC) in Berkeley, California. The research was performed at the Ames National Laboratory, which is operated for the U.S. DOE by Iowa State University under Contract No. DE-AC02-07CH11358. 

\section*{Author Contributions}
T.J. and Y.-X.Y. conceived and designed the project. T.J. performed the calculations. T.J., Y.-X.Y., and J.W. analyzed the results and wrote the manuscript. All authors discussed the results and commented on the manuscript.

\section*{Competing Interests}
The authors declare no competing financial or non-financial interests.

\renewcommand{\bibsection}{\section*{References}}
\bibliography{refabbrev, ref}
% =========================
% Tables
% =========================

% =========================
% Figures
% =========================

% ============================================================
% Supplementary Information
% ============================================================

\clearpage
\onecolumngrid

% Reset supplementary counters
\setcounter{equation}{0}
\setcounter{figure}{0}
\setcounter{table}{0}
\setcounter{section}{0}

\renewcommand{\theequation}{S\arabic{equation}}
\renewcommand{\thefigure}{S\arabic{figure}}
\renewcommand{\thetable}{S\arabic{table}}

\vspace*{1em}

\begin{center}
{\Large\bfseries Supplementary Information}

\vspace{0.8em}

{\large
CDW Gap Collapse and Weyl State Restoration in
Ta$_2$Se$_8$I via Coherent Phonons:\\
A First-Principles Study
}

\vspace{0.8em}

Tao Jiang*, Jigang Wang, and Yong-Xin Yao
\end{center}

\vspace{1.5em}

\section*{Supplementary Notes}

% =====================================================================
% CONTENTS: Clickable blue table of contents; chapter numbers and page numbers are automatically generated.
% =====================================================================
\noindent{\bf CONTENTS}
\vspace{0.5em}

\tocentry{sec:pathgap}{Direct-gap collapse on different high-symmetry paths}
\tocentry{sec:weights}{Element-resolved weights of the $\Gamma\text{--}\mathrm{Z}$ bands}
\tocentry{sec:element}{Element-resolved contribution of the $A(18)$ mode to gap suppression}
\tocentry{sec:modes}{Zone-center phonon modes of (TaSe$_4$)$_2$I in the CDW phase}
\tocentry{sec:couplingall}{Anharmonic coupling between IR-active phonons and the CDW amplitude mode $A(18)$}
\tocentry{sec:kpoint}{K-point convergence of the $\Gamma\text{--}\mathrm{Z}$ direct gap}
\tocentry{sec:wannier}{Wannier and Weyl-node validation}
\tocentry{sec:fluence}{Estimate of coherent-phonon energy and pump fluence}
\tocentry{sec:thz}{Mode effective charge and THz field strength for driving the $B_3(7)$ mode}
\tocentry{sec:rectified}{Estimate of the rectified $A(18)$ displacement induced by the $B_3(7)$ mode}
\tocentry{sec:dispersion}{Full phonon dispersion and dynamical stability}
\tocentry{sec:pbesol}{PBEsol validation of phonon-driven Weyl-state restoration}

\vspace{2em}
\hrule
\vspace{2em}

\section{Direct-gap collapse on different high-symmetry paths}
\label{sec:pathgap}
\begin{figure}[t]
  \centering
  \includegraphics[width=3.5in]{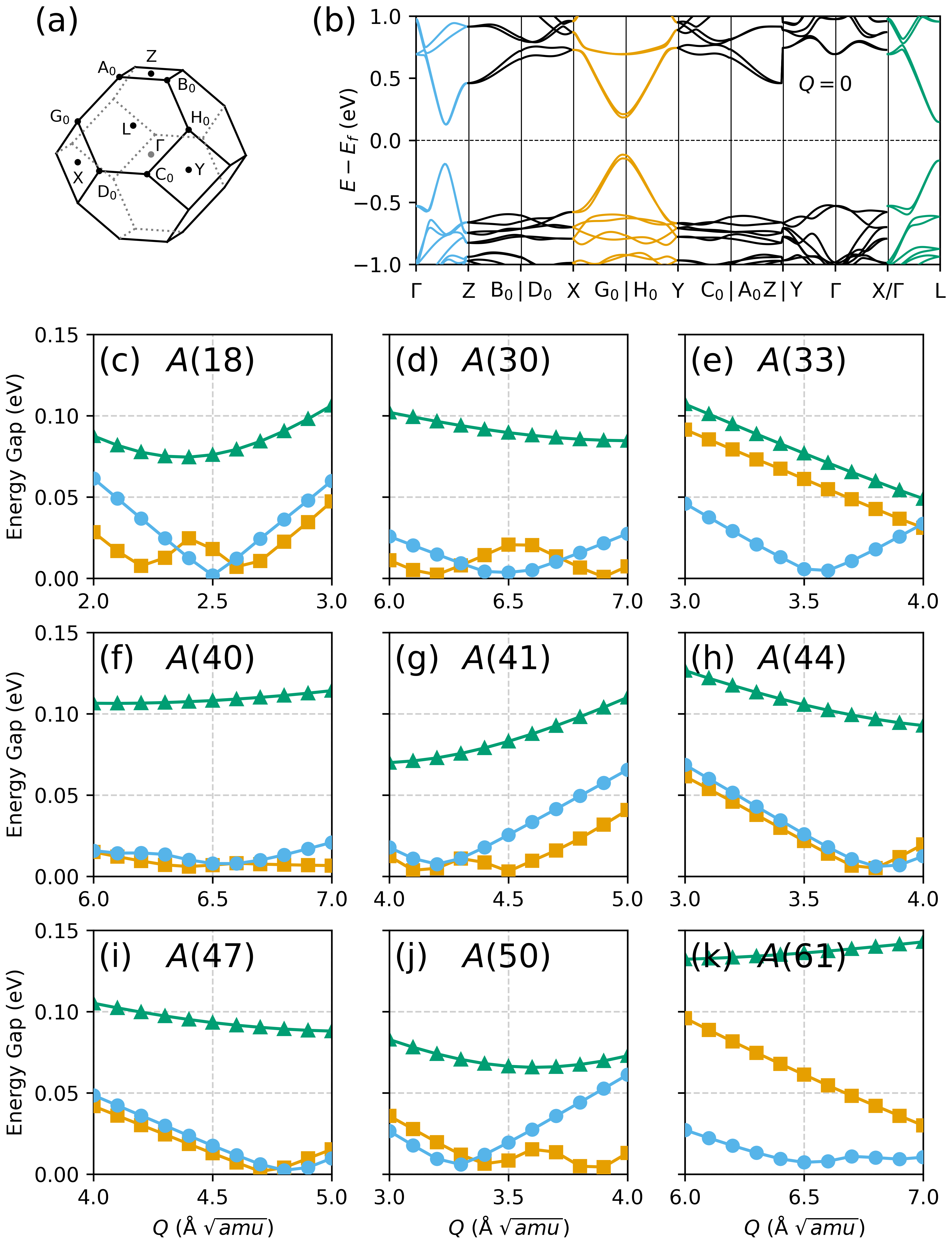}
  \caption{Direct-gap response on different high-symmetry paths under frozen-phonon distortions in the $F222$ CDW phase of $(\mathrm{TaSe}_4)_2\mathrm{I}$. (a) Brillouin zone and the high-symmetry points used in this section. (b) Electronic band structure at equilibrium ($Q=0$) along the selected high-symmetry paths; the colored segments highlight the three paths used for direct-gap tracking: $\Gamma\text{--}\mathrm{Z}$ (blue), $\mathrm{X}\text{--}\mathrm{G}_0\!\mid\!\mathrm{H}_0\text{--}\mathrm{Y}$ (yellow), and $\Gamma\text{--}\mathrm{L}$ (green). (c)--(k) Minimum direct band gap as a function of the normal-mode amplitude $Q$ for $A(18)$, $A(30)$, $A(33)$, $A(40)$, $A(41)$, $A(44)$, $A(47)$, $A(50)$, and $A(61)$, respectively, comparing the three paths on the same energy scale and with identical electronic-structure settings as in the main text.
  }
  \label{fig:pathgap}
\end{figure}

To assess whether the minimum direct $\Gamma\text{--}\mathrm{Z}$ gap provides a reliable operational descriptor of CDW-gap suppression in the main text, we compared the evolution of the minimum direct band gap along several representative high-symmetry paths, including $\mathrm{\Gamma}\text{--}\mathrm{Z}$, $\mathrm{X}\text{--}\mathrm{G}_{0}\!\mid\!\mathrm{H}_{0}\text{--}\mathrm{Y}$, and $\mathrm{\Gamma}\text{--}\mathrm{L}$, as a function of the normal-mode amplitude $Q$ for the same set of frozen-phonon distortions. For each distorted structure, we extracted the minimum direct gap along each path by scanning the band energies on the same $\mathrm{k}$-point sampling and using identical electronic-structure settings as in the main text. As shown in Figure~\ref{fig:pathgap}, the direct gaps along $\mathrm{\Gamma}\text{--}\mathrm{Z}$ and $\mathrm{X}\text{--}\mathrm{G}_{0}\!\mid\!\mathrm{H}_{0}\text{--}\mathrm{Y}$ are, in general, suppressed much more rapidly with increasing $Q$ than the gap along $\mathrm{\Gamma}\text{--}\mathrm{L}$. For most of the principal Raman modes considered here, the $\mathrm{\Gamma}\text{--}\mathrm{Z}$ gap collapses at a rate comparable to or faster than the $\mathrm{X}\text{--}\mathrm{G}_{0}\!\mid\!\mathrm{H}_{0}\text{--}\mathrm{Y}$ gap. Three minor exceptions are the CDW amplitude mode $A(18)$, mode $A(30)$ and mode $A(40)$, for which $\mathrm{X}\text{--}\mathrm{G}_{0}\!\mid\!\mathrm{H}_{0}\text{--}\mathrm{Y}$ decreases marginally faster than $\mathrm{\Gamma}\text{--}\mathrm{Z}$ over a limited range of small-to-intermediate $Q$. Any deviations from this trend are marginal and restricted to a limited $Q$ window, and therefore do not affect the conclusions drawn below. In particular, for $A(18)$ the minimum gap along $\mathrm{X}\text{--}\mathrm{G}_{0}\!\mid\!\mathrm{H}_{0}\text{--}\mathrm{Y}$ reaches the near-closure regime at $Q\simeq 2.2~\mathrm{\AA}\sqrt{\mathrm{amu}}$, slightly earlier than along $\mathrm{\Gamma}\text{--}\mathrm{Z}$ (near $Q\simeq 2.5~\mathrm{\AA}\sqrt{\mathrm{amu}}$). This modest difference between the two paths does not determine either the location or the onset of the Weyl nodes, both of which are established independently by the full-Brillouin-zone search. Indeed, Weyl nodes are already present at \(Q=2.0~\mathrm{\AA}\sqrt{\mathrm{amu}}\), before either tracked path reaches its gap-minimizing amplitude. Importantly, this subtle path-dependent difference does not affect our main conclusions: the emergence of Weyl nodes is established by Brillouin-zone--wide searches, and Weyl points already appear at $Q = 2.0~\mathrm{\AA}\sqrt{\mathrm{amu}}$ even before the $\mathrm{\Gamma}\text{--}\mathrm{Z}$ direct gap reaches its near-closure minimum. A similar, weak preference of $\mathrm{X}\text{--}\mathrm{G}_{0}\!\mid\!\mathrm{H}_{0}\text{--}\mathrm{Y}$ is observed for $A(30)$ and $A(40)$. In contrast, the $\mathrm{\Gamma}\text{--}\mathrm{L}$ gap decreases more slowly and typically remains finite over the same distortion range, demonstrating a reduced sensitivity to the phonon-driven lattice modulation. Taken together, these comparisons support the use of the minimum direct $\mathrm{\Gamma}\text{--}\mathrm{Z}$ gap as a robust operational descriptor for the principal Raman modes. This trend is consistent with the quasi-one-dimensional nature of $(\mathrm{TaSe}_4)_2\mathrm{I}$, where the CDW distortion predominantly modulates electronic states dispersing along the chain direction (reciprocal $c^\ast$), making $\mathrm{\Gamma}\text{--}\mathrm{Z}$ particularly responsive to the CDW-induced gap opening and its phonon-driven collapse. Therefore, we use the minimum direct $\mathrm{\Gamma}\text{--}\mathrm{Z}$ gap as the primary operational descriptor for comparing phonon-induced CDW-gap suppression in the present analysis, while the actual Weyl-semimetallic state is identified independently by the Brillouin-zone--wide Weyl-node search.

\section{Element-resolved weights of the $\Gamma\text{--}\mathrm{Z}$ bands}
\label{sec:weights}
\begin{figure}
    \centering
    \includegraphics[width=3.5in]{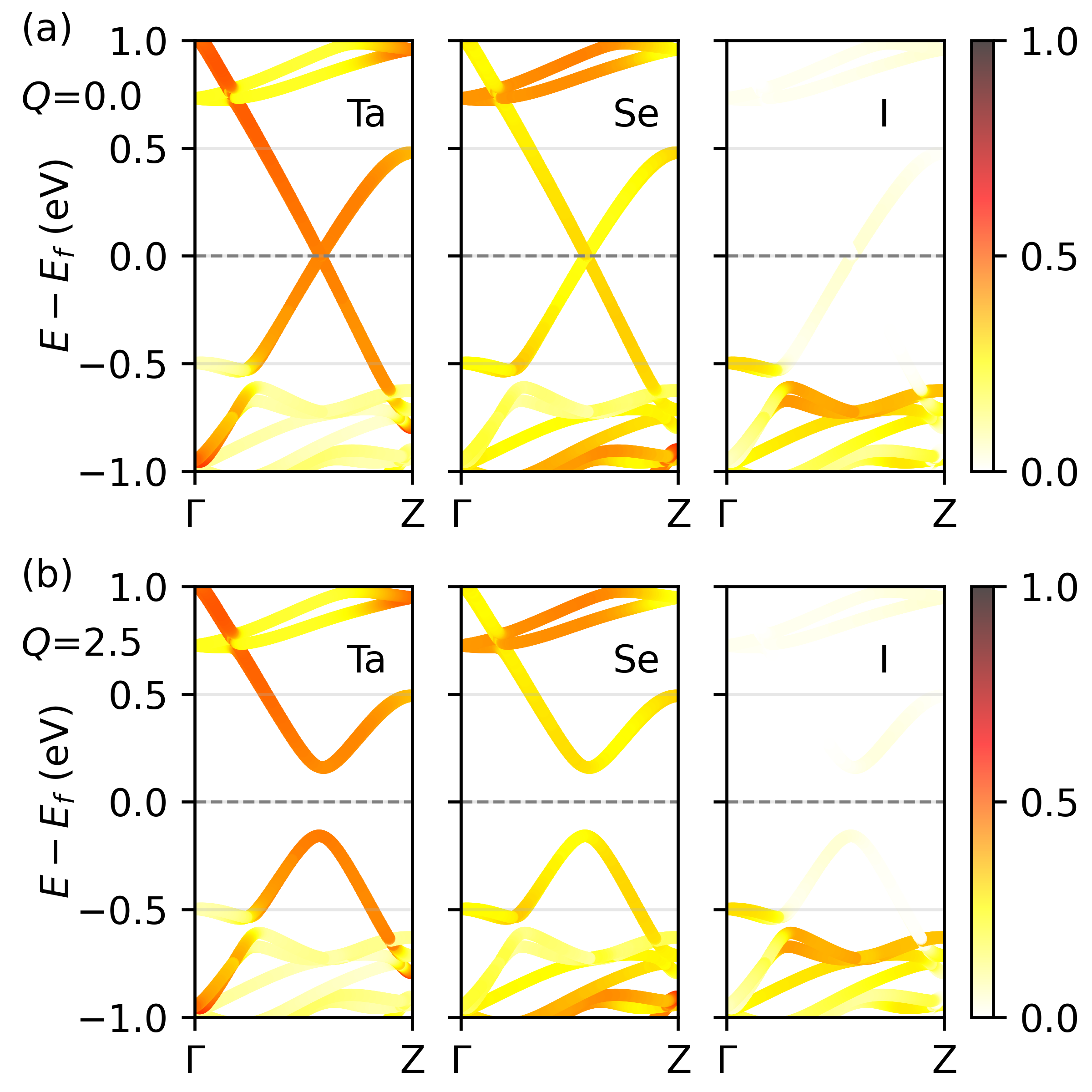}
    \caption{Element-resolved projected band weights along $\Gamma\text{--}\mathrm{Z}$ for (TaSe$_4$)$_2$I. (a) Equilibrium CDW structure ($Q=0$). (b) \(A(18)\)-distorted structure at $Q=2.5~\text{\AA}\sqrt{\text{amu}}$. For each structure, the three panels show the Ta-, Se-, and I-projected weights, respectively. The color scale indicates the normalized projected weight (0--1) of the corresponding element for each Kohn--Sham band state along $\Gamma\text{--}\mathrm{Z}$.}
\label{fig:weights}
\end{figure}

To characterize the elemental composition of the near-gap electronic states, we projected the band weights onto Ta, Se, and I along the $\Gamma\text{--}\mathrm{Z}$ path for the equilibrium CDW structure ($Q=0$) and for the \(A(18)\)-distorted structure at $Q=2.5~\text{\AA}\sqrt{\text{amu}}$ (Figure~\ref{fig:weights}). The color intensity encodes the normalized projected weight of a given element for each Kohn–Sham band state along $\Gamma\text{--}\mathrm{Z}$, where values closer to 1 indicate a stronger contribution from that element. 
For both $Q=0$ and $Q=2.5$, the states forming the valence-band maximum and conduction-band minimum are dominated by Ta weight with a secondary Se contribution, while the I contribution remains negligible near the Fermi level. This element-resolved analysis supports the conclusion that the \(A(18)\)-induced gap suppression is primarily governed by distortions that strongly affect the Ta-derived near-gap bands, with Se providing a weaker but non-negligible modulation.

\section{Element-resolved contribution of the \(A(18)\) mode to gap suppression}
\label{sec:element}
\begin{figure}
    \centering
    \includegraphics[width=3.5in]{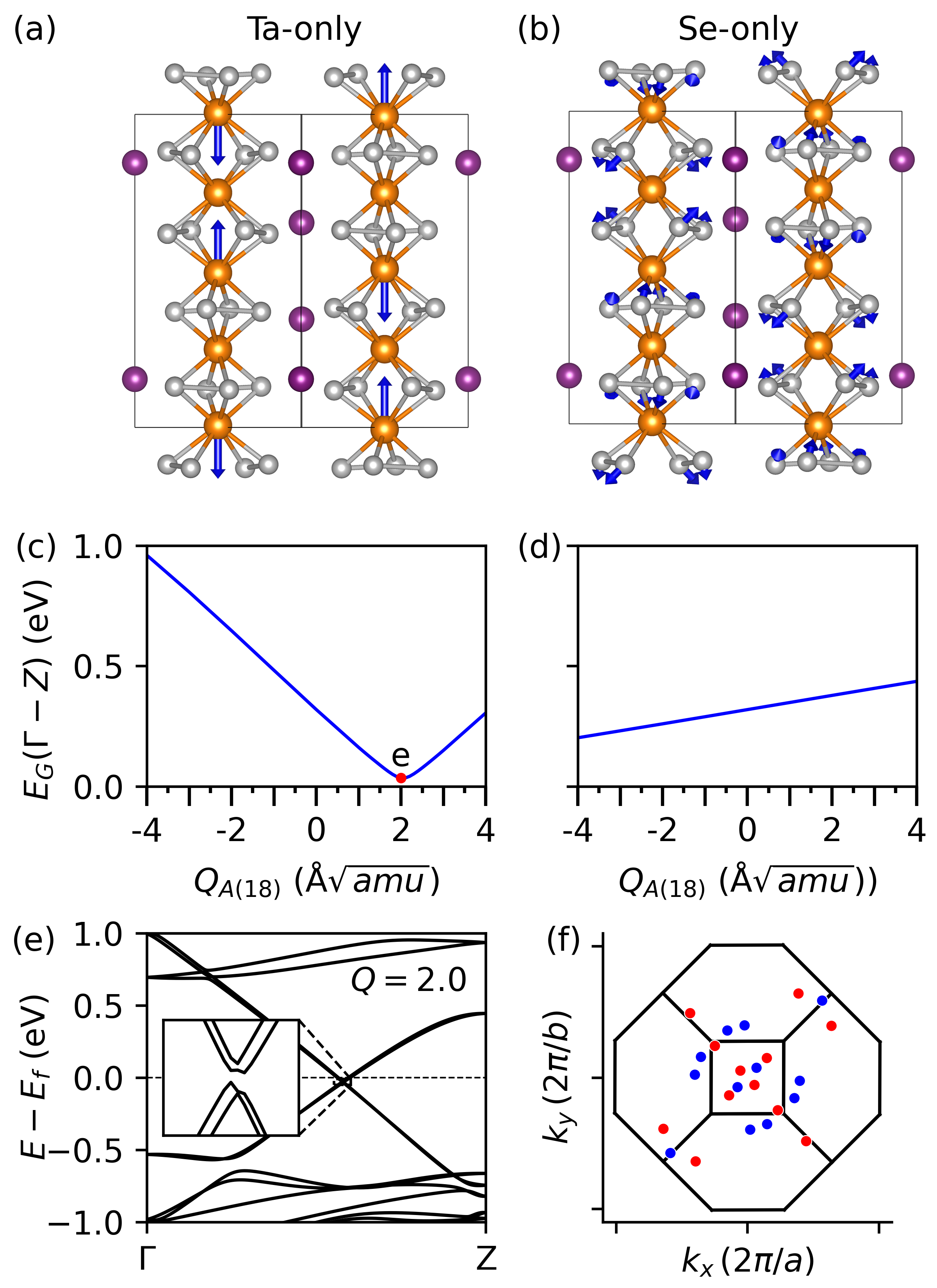}
    \caption{Element-selective decomposition of the CDW amplitude mode \(A(18)\) and its distinct impacts on the $\Gamma\text{--}\mathrm{Z}$ gap and Weyl topology. (a) Ta-only distortion constructed by keeping only the Ta atomic displacements from the full \(A(18)\) eigenvector (Se and I atoms fixed). (b) Se-only distortion constructed by keeping only the Se displacements (Ta and I fixed). (c),(d) Evolution of the minimum direct gap along $\Gamma\text{--}\mathrm{Z}$, $E_G(\Gamma\text{--}\mathrm{Z})$, as a function of the normal-mode coordinate $Q_{A(18)}$ for the Ta-only and Se-only distortions in (a) and (b), respectively. (e) Electronic band structure along $\Gamma\text{--}\mathrm{Z}$ at the Ta-only distortion marked in (c), where the minimum direct $\Gamma\text{--}\mathrm{Z}$ gap is suppressed to 4.1~meV and the residual opening is resolved in the magnified inset. The inset shows a magnified view in the energy window $E-E_F\in[-0.05,-0.01]$~eV. (f) Corresponding Weyl-node distribution obtained from a Brillouin-zone--wide search for the Ta-only distorted structure in (e), shown as a $k_x$--$k_y$ projection; red and blue spheres denote Weyl nodes with positive and negative chiral charge, respectively. Blue arrows indicate the atomic displacement directions. Orange, gray, and purple spheres denote Ta, Se, and I atoms, respectively.}
    \label{fig:element}
\end{figure}

To further clarify the microscopic origin of the \(A(18)\)-driven gap suppression, we analyzed the element-resolved character of the \(A(18)\) lattice distortion. We decomposed the \(A(18)\) eigenvector into element-resolved components by constructing ``Ta-only'' and ``Se-only'' frozen-phonon distortions (Figure~\ref{fig:element}). 
In the Ta-only distortion, only Ta atoms are displaced following the \(A(18)\) eigenvector direction while Se and I atoms are fixed [Figure~\ref{fig:element}(a)]; conversely, the Se-only distortion displaces only Se atoms while Ta and I atoms remain fixed [Figure~\ref{fig:element}(b)]. 
In both cases we keep the same normalization of the normal-mode coordinate $Q_{A(18)}$ as in the full-mode distortion to enable a direct comparison.

Figures~\ref{fig:element}(c) and \ref{fig:element}(d) show the evolution of the minimum direct gap along $\Gamma\text{--}\mathrm{Z}$, $E_G(\Gamma\text{--}\mathrm{Z})$, as a function of $Q_{A(18)}$ for the Ta-only and Se-only distortions, respectively. 
The Ta-only distortion produces a pronounced suppression of the $\Gamma\text{--}\mathrm{Z}$ gap and already reduces it to the meV scale at $Q_{A(18)}= 2.0~\text{\AA}\sqrt{\text{amu}}$, which is smaller than the amplitude required for the full \(A(18)\) mode ($Q_{A(18)}=2.5~\text{\AA}\sqrt{\text{amu}}$). 
By contrast, the Se-only distortion drives the gap in the opposite direction and yields a weak \emph{increase} of $E_G(\Gamma\text{--}\mathrm{Z})$ for $Q>0$. 
This opposite trend demonstrates that the Se contribution enters with the opposite sign in the gap modulation and thus partially counteracts the Ta-driven gap suppression in the full mode, consistent with the Ta-only and Se-only frozen-phonon tests. Figure~\ref{fig:element}(e) shows the corresponding $\Gamma\text{--}\mathrm{Z}$ band structure for the Ta-only distortion at $Q_{A(18)}=2.0~\text{\AA}\sqrt{\text{amu}}$ (marked in Figure~\ref{fig:element}(c)), where the minimum direct gap is reduced to the meV scale; the magnified inset resolves the residual gap near $E_F$.

A Brillouin-zone--wide Weyl-node search based on the Wannier Hamiltonian identifies Weyl nodes at generic \(k\) points for this Ta-only distorted structure, and their $k_x$--$k_y$ projection is shown in Figure~\ref{fig:element}(f). 
Taken together, these results indicate that the \(A(18)\) mode couples to the gap through opposite-signed element-resolved contributions: Ta motion provides the leading gap-suppression channel, whereas Se motion provides a weaker gap-opening modulation, so that the net $Q_\mathrm{c}$ reflects a competition between Ta and Se displacements rather than Ta--Ta geometry alone.

\section{Zone-Center Phonon Modes of (TaSe$_4$)$_2$I in the CDW Phase}
\label{sec:modes}
We list all zone-center phonon modes of (TaSe$_4$)$_2$I in its CDW phase (space group $F$222), including their vibrational frequencies, irreducible representations, and optical activity. This table provides the mode indices used throughout the main text, including the Raman-active CDW amplitude mode \(A(18)\) at 2.52~THz and the additional symmetry-preserving Raman \(A\) modes (\(A(30)\), \(A(33)\), \(A(40)\), \(A(41)\), \(A(44)\), \(A(47)\), \(A(50)\), and \(A(61)\)) that are dominated by Se vibrations. We also report the space group retained after displacing atoms along each phonon eigenvector (SG$_\mathrm{disp}$), which is used to distinguish symmetry-preserving and symmetry-breaking distortions.

\begin{table*}[t]
\caption{Zone-center phonon modes of (TaSe$_4$)$_2$I in the CDW phase ($F$222), including vibrational frequencies (THz), irreducible representations (Irrep), optical activity, and the space group retained after displacing atoms along each phonon eigenvector (SG$_\mathrm{disp}$).}
\label{Table:A1}
\centering
\begin{ruledtabular}
\begin{tabular}{r r c l l | r r c l l | r r c l l}
Idx & Freq & Irrep & Activity & SG$_\mathrm{disp}$ &
Idx & Freq & Irrep & Activity & SG$_\mathrm{disp}$ &
Idx & Freq & Irrep & Activity & SG$_\mathrm{disp}$ \\
\hline
1  & 0.00 & \(B_2\) & Acoustic & $F$222 & 23 & 3.10 & \(B_1\) & IR,Raman & $C$2 & 45 & 5.10 & \(B_2\) & IR,Raman & $C$2 \\
2  & 0.00 & \(B_1\)  & Acoustic & $F$222 & 24 & 3.21 & \(B_2\) & IR,Raman & $C$2 & 46 & 5.25 & \(B_3\) & IR,Raman & $C$2 \\
3  & 0.00 & \(B_3\) & Acoustic & $F$222 & 25 & 3.25 & \(B_1\) & IR,Raman & $C$2 & 47 & 5.34 & \(A\)   & Raman    & $F$222 \\
4  & 1.01 & \(B_1\) & IR,Raman & $C$2 & 26 & 3.71 & \(B_3\) & IR,Raman & $C$2 & 48 & 5.48 & \(B_2\) & IR,Raman & $C$2 \\
5  & 1.04 & \(B_3\) & IR,Raman & $C$2 & 27 & 3.95 & \(B_2\) & IR,Raman & $C$2 & 49 & 5.48 & \(B_1\) & IR,Raman & $C$2 \\
6  & 1.06 & \(B_2\) & IR,Raman & $C$2 & 28 & 3.98 & \(B_1\) & IR,Raman & $C$2 & 50 & 5.51 & \(A\)   & Raman    & $F$222 \\
7  & 1.15 & \(B_3\) & IR,Raman & $C$2 & 29 & 4.01 & \(B_3\) & IR,Raman & $C$2 & 51 & 5.65 & \(B_2\) & IR,Raman & $C$2 \\
8  & 1.27 & \(B_2\) & IR,Raman & $C$2 & 30 & 4.07 & \(A\)   & Raman    & $F$222 & 52 & 5.67 & \(B_3\) & IR,Raman & $C$2 \\
9  & 1.30 & \(B_1\) & IR,Raman & $C$2 & 31 & 4.16 & \(B_1\) & IR,Raman & $C$2 & 53 & 5.86 & \(B_3\) & IR,Raman & $C$2 \\
10 & 1.55 & \(B_2\) & IR,Raman & $C$2 & 32 & 4.22 & \(B_2\) & IR,Raman & $C$2 & 54 & 5.87 & \(B_1\) & IR,Raman & $C$2 \\
11 & 1.58 & \(B_1\) & IR,Raman & $C$2 & 33 & 4.23 & \(A\)   & Raman    & $F$222 & 55 & 6.11 & \(B_2\) & IR,Raman & $C$2 \\
12 & 1.60 & \(B_1\) & IR,Raman & $C$2 & 34 & 4.30 & \(B_3\) & IR,Raman & $C$2 & 56 & 6.29 & \(B_1\) & IR,Raman & $C$2 \\
13 & 1.62 & \(B_2\) & IR,Raman & $C$2 & 35 & 4.44 & \(A\)   & Raman    & $F$222 & 57 & 6.66 & \(B_1\) & IR,Raman & $C$2 \\
14 & 1.67 & \(B_3\) & IR,Raman & $C$2 & 36 & 4.48 & \(B_2\) & IR,Raman & $C$2 & 58 & 6.75 & \(B_2\) & IR,Raman & $C$2 \\
15 & 1.82 & \(A\)   & Raman    & $F$222 & 37 & 4.50 & \(B_1\) & IR,Raman & $C$2 & 59 & 7.76 & \(B_2\) & IR,Raman & $C$2 \\
16 & 1.91 & \(A\)   & Raman    & $F$222 & 38 & 4.50 & \(B_3\) & IR,Raman & $C$2 & 60 & 7.81 & \(B_1\) & IR,Raman & $C$2 \\
17 & 2.01 & \(B_3\) & IR,Raman & $C$2 & 39 & 4.75 & \(B_3\) & IR,Raman & $C$2 & 61 & 7.84 & \(A\)   & Raman    & $F$222 \\
18 & 2.52 & \(A\)   & Raman    & $F$222 & 40 & 4.82 & \(A\)   & Raman    & $F$222 & 62 & 7.84 & \(B_3\) & IR,Raman & $C$2 \\
19 & 2.71 & \(B_1\) & IR,Raman & $C$2 & 41 & 4.84 & \(A\)   & Raman    & $F$222 & 63 & 7.91 & \(A\)   & Raman    & $F$222 \\
20 & 2.73 & \(A\)   & Raman    & $F$222 & 42 & 4.91 & \(B_3\) & IR,Raman & $C$2 & 64 & 8.05 & \(B_3\) & IR,Raman & $C$2 \\
21 & 2.79 & \(B_2\) & IR,Raman & $C$2 & 43 & 4.97 & \(B_1\) & IR,Raman & $C$2 & 65 & 8.13 & \(B_2\) & IR,Raman & $C$2 \\
22 & 3.07 & \(B_2\) & IR,Raman & $C$2 & 44 & 5.08 & \(A\)   & Raman    & $F$222 & 66 & 8.17 & \(B_1\) & IR,Raman & $C$2 \\
\end{tabular}
\end{ruledtabular}
\end{table*}

\section{Anharmonic Coupling Between IR-Active Phonons and the CDW Amplitude Mode $A(18)$}
\label{sec:couplingall}
This section summarizes the anharmonic coupling coefficients between \emph{all} infrared-active phonon modes and the Raman-active CDW amplitude mode \(A(18)\), extracted from the fourth-order nonlinear phonon potential in Eq.~(2) of the main text. The coefficients are obtained by fitting the two-dimensional energy surface \(V(Q_{\mathrm{IR}},Q_{\mathrm{R}})\) computed from frozen-phonon DFT calculations (energy in meV per atom). For clarity, we list only symmetry-allowed, nonzero terms; coefficients associated with odd powers of \(Q_{\mathrm{IR}}\) vanish for all IR \(B\) modes. Among all IR modes, mode 7 shows a sizable cubic and the strongest quartic IR--Raman coupling and is referred to as \(B_3(7)\) in the main text.

\begin{table*}[t]
\caption{Complete list of anharmonic coupling coefficients between \emph{all} IR-active phonon modes and the Raman-active CDW amplitude mode \(A(18)\) in (TaSe$_4$)$_2$I. Listed are the mode indices and selected coefficients ($c_2$, $c_{21}$, $d_1$, $d_2$, $d_{22}$) from the nonlinear energy expansion. Coefficients not included in this table are $c_1$, $c_{12}$, $d_{13}$, and $d_{31}$, as they are zero for all modes.}
\label{Table:A2}
\centering
\begin{ruledtabular}
\begin{tabular}{r r r r r r | r r r r r r | r r r r r r}
Idx & $c_2$ & $c_{21}$ & $d_1$ & $d_2$ & $d_{22}$ &
Idx & $c_2$ & $c_{21}$ & $d_1$ & $d_2$ & $d_{22}$ &
Idx & $c_2$ & $c_{21}$ & $d_1$ & $d_2$ & $d_{22}$ \\
\hline
4  & -0.191 &  0.009 & 0.026 & -0.020 & -0.001 &
25 & -0.190 & -0.001 & 0.031 & -0.019 & -0.003 &
49 & -0.189 &  0.012 & 0.030 & -0.018 & -0.006 \\

5  & -0.190 &  0.004 & 0.007 & -0.019 & -0.001 &
26 & -0.191 &  0.002 & 0.032 & -0.019 & -0.002 &
51 & -0.190 &  0.037 & 0.038 & -0.019 & -0.002 \\

6  & -0.190 &  0.005 & 0.034 & -0.019 & -0.002 &
27 & -0.191 &  0.015 & 0.025 & -0.019 & -0.001 &
52 & -0.190 & -0.037 & 0.032 & -0.018 & -0.003 \\

7  & -0.180 & -0.080 & -0.002 & -0.019 &  0.009 &
28 & -0.190 &  0.020 & 0.028 & -0.019 & -0.002 &
53 & -0.189 &  0.005 & 0.036 & -0.019 & -0.002 \\

8  & -0.190 &  0.005 & 0.020 & -0.019 & -0.002 &
29 & -0.187 & -0.026 & 0.017 & -0.019 & -0.001 &
54 & -0.190 & -0.068 & 0.034 & -0.019 & -0.005 \\

9  & -0.190 &  0.003 & 0.034 & -0.019 & -0.002 &
31 & -0.190 &  0.029 & 0.027 & -0.019 & -0.001 &
55 & -0.191 &  0.031 & 0.036 & -0.019 & -0.005 \\

10 & -0.191 &  0.012 & 0.033 & -0.019 & -0.002 &
32 & -0.191 & -0.014 & 0.028 & -0.019 & -0.004 &
56 & -0.190 & -0.022 & 0.040 & -0.019 & -0.001 \\

11 & -0.191 &  0.005 & 0.026 & -0.020 & -0.002 &
34 & -0.189 &  0.077 & 0.036 & -0.019 &  0.000 &
57 & -0.189 &  0.018 & 0.032 & -0.019 & -0.001 \\

12 & -0.191 &  0.006 & 0.020 & -0.020 & -0.001 &
36 & -0.190 &  0.032 & 0.028 & -0.019 & -0.002 &
58 & -0.190 & -0.024 & 0.044 & -0.019 & -0.004 \\

13 & -0.190 &  0.004 & 0.024 & -0.020 & -0.002 &
37 & -0.191 & -0.011 & 0.030 & -0.019 & -0.002 &
59 & -0.190 & -0.005 & 0.085 & -0.019 & -0.001 \\

14 & -0.190 &  0.003 & 0.033 & -0.019 & -0.001 &
38 & -0.190 &  0.013 & 0.039 & -0.019 &  0.000 &
60 & -0.190 & -0.014 & 0.081 & -0.019 & -0.002 \\

17 & -0.189 & -0.001 & 0.032 & -0.019 & -0.001 &
39 & -0.188 & -0.014 & 0.024 & -0.019 & -0.002 &
62 & -0.189 & -0.005 & 0.094 & -0.019 & -0.001 \\

19 & -0.190 &  0.013 & 0.042 & -0.019 & -0.002 &
42 & -0.191 & -0.085 & 0.040 & -0.018 & -0.005 &
64 & -0.189 & -0.028 & 0.092 & -0.019 &  0.001 \\

21 & -0.190 &  0.001 & 0.031 & -0.019 & -0.002 &
43 & -0.191 &  0.042 & 0.030 & -0.019 & -0.003 &
65 & -0.189 & -0.013 & 0.069 & -0.019 & -0.002 \\

22 & -0.190 &  0.019 & 0.031 & -0.019 & -0.002 &
45 & -0.190 &  0.026 & 0.042 & -0.019 &  0.002 &
66 & -0.189 & -0.027 & 0.060 & -0.019 & -0.002 \\

23 & -0.191 & -0.003 & 0.032 & -0.019 & -0.001 &
46 & -0.186 & -0.042 & 0.027 & -0.019 &  0.002 &
   &        &        &       &        &        \\

24 & -0.190 &  0.013 & 0.027 & -0.019 & -0.002 &
48 & -0.189 & -0.008 & 0.036 & -0.019 &  0.002 &
   &        &        &       &        &        \\
\end{tabular}
\end{ruledtabular}
\end{table*}

% =====================================================================
% ---------------------------------------------------------------------
% I. K-point convergence
% ---------------------------------------------------------------------
% =====================================================================

\section{K-Point Convergence of the $\Gamma\text{--}Z$ Direct Gap}
\label{sec:kpoint}

Because the phonon-driven CDW-gap suppression discussed in the main text reaches the meV scale near \(Q=2.5~\mathrm{\AA}\sqrt{\mathrm{amu}}\), we performed explicit self-consistent k-point convergence tests for the minimum direct gap along the \(\Gamma\text{--}Z\) direction. The tests were carried out for both the equilibrium \(F222\) CDW structure and the \(A(18)\)-distorted structure at \(Q=2.5~\mathrm{\AA}\sqrt{\mathrm{amu}}\), using \(4\times4\times4\), \(6\times6\times6\), and \(8\times8\times8\) meshes. The same SOC-included PBE functional, plane-wave cutoff, electronic convergence settings, and band-path sampling were used in all calculations; only the self-consistent k-point mesh was varied.

For the equilibrium structure, the minimum direct gap changes from \(316.10\)~meV with the \(4\times4\times4\) mesh to \(317.20\)~meV with both the \(6\times6\times6\) and \(8\times8\times8\) meshes. For the \(A(18)\)-distorted structure, the corresponding residual gaps are \(0.97\), \(0.57\), and \(0.46\)~meV, respectively. In particular, increasing the sampling from \(6\times6\times6\) to \(8\times8\times8\) changes the residual gap by only \(0.11\)~meV.

\begin{table}[htbp]
\caption{
K-point convergence of the minimum direct gap along the
\(\Gamma\text{--}Z\) direction for the equilibrium \(F222\) CDW
structure and the \(A(18)\)-distorted structure at
\(Q=2.5~\mathrm{\AA}\sqrt{\mathrm{amu}}\). The same SOC-included PBE
settings and band-path sampling were used for all calculations.
}
\label{tab:S_kpoint_convergence_gap}
\centering
\begin{ruledtabular}
\begin{tabular}{c c c}
SCF k-point mesh
& \(E_{\mathrm{gap}}^{\min}(Q=0)\)
& \(E_{\mathrm{gap}}^{\min}[A(18),Q=2.5]\) \\
& (meV) & (meV) \\
\hline
\(4\times4\times4\) & 316.10 & 0.97 \\
\(6\times6\times6\) & 317.20 & 0.57 \\
\(8\times8\times8\) & 317.20 & 0.46 \\
\end{tabular}
\end{ruledtabular}
\end{table}

These convergence tests demonstrate that the equilibrium CDW gap is well converged and that the strongly suppressed residual gap remains robustly in the sub-meV range with respect to the self-consistent k-point mesh. In particular, increasing the sampling from \(6\times6\times6\) to \(8\times8\times8\) changes the residual gap by only \(0.11\)~meV. The strong suppression of the \(\Gamma\text{--}Z\) direct gap by the symmetry-preserving Raman-active \(A(18)\) mode is therefore robust with respect to k-point sampling. We consequently use the \(6\times6\times6\) mesh for the production calculations reported in the main text.

% =====================================================================
% ---------------------------------------------------------------------
% II. Wannierization and Weyl-node search
% ---------------------------------------------------------------------
% =====================================================================

\section{Wannier and Weyl-Node Validation}
\label{sec:wannier}

To ensure that the Weyl nodes discussed in the main text are not
artifacts of band interpolation or of an analysis restricted to
high-symmetry paths, we performed an explicit Wannier-based search
over the full three-dimensional Brillouin zone. For each
phonon-distorted structure analyzed in the main text, a
spin--orbit-coupled Wannier tight-binding Hamiltonian was constructed
from the SOC-included PBE calculation using \textsc{Wannier90} and
subsequently analyzed with \textsc{WannierTools}. All construction
and search parameters are collected in
Table~\ref{tab:S_wannier_settings}. The number of Wannier functions,
\texttt{num\_wann}=148, corresponds to the 74 orbital projections$4\times5_{\mathrm{Ta}\,d}+16\times3_{\mathrm{Se}\,p}+2\times3_{\mathrm{I}\,p}=74$,doubled by the spinor degree of freedom. The Fermi level and energy
reference were kept consistent with the corresponding DFT calculation
for each structure.

\begin{figure}[htbp]
\centering
\includegraphics[width=\textwidth]{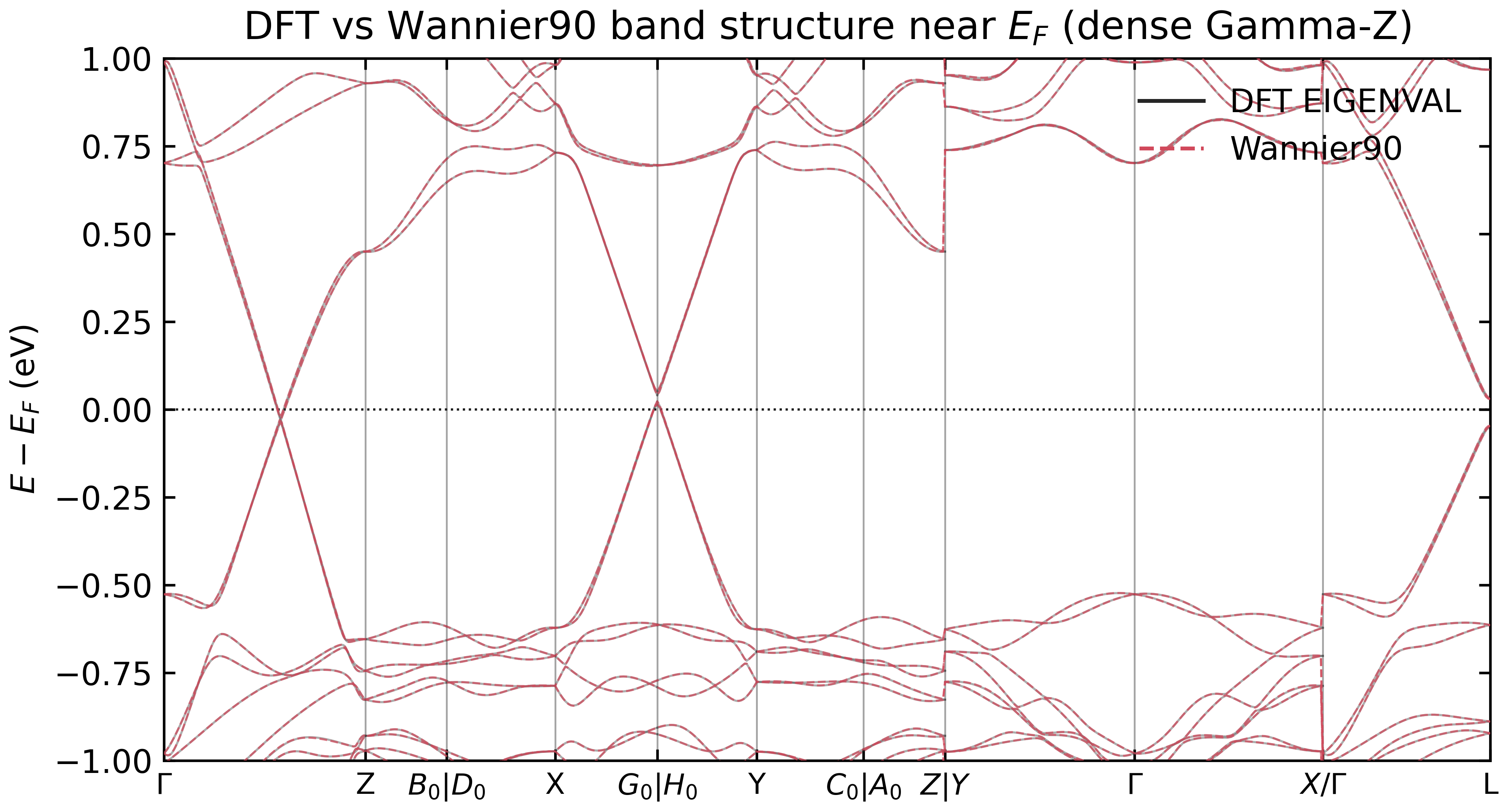}
\caption{
{\bf Accuracy of the Wannier interpolation used for the Weyl-node
search.}
Comparison of the SOC-included DFT eigenvalues (black solid lines) with the
Wannier-interpolated bands (red dashed lines) along the complete
high-symmetry path for the \(A(18)\)-distorted structure at
\(Q=2.5~\mathrm{\AA}\sqrt{\mathrm{amu}}\). This is one of the
structures for which Weyl nodes are reported in the main text. The
comparison is shown over the energy window
\(E-E_F\in[-1,1]\)~eV. The close agreement in the band ordering and
dispersion, particularly for the low-energy bands involved in the
CDW-gap suppression, supports the use of the Wannier Hamiltonian in
the full-Brillouin-zone topological analysis.
}
\label{fig:S_wannier_bands}
\end{figure}

The fidelity of the Wannier interpolation was verified by direct
comparison with the underlying SOC-included DFT band structure for
the \(A(18)\)-distorted structure at
\(Q=2.5~\mathrm{\AA}\sqrt{\mathrm{amu}}\), one of the structures for
which Weyl nodes are reported in the main text. As shown in
Figure~\ref{fig:S_wannier_bands}, the Wannier-interpolated bands closely
reproduce the DFT band ordering and dispersion along the complete
high-symmetry path over the energy window
\(E-E_F\in[-1,1]\)~eV. In particular, the low-energy bands involved
in the CDW-gap suppression are well reproduced, supporting the use
of the Wannier Hamiltonian for the subsequent full-Brillouin-zone
topological analysis.

Because the identified Weyl nodes occur at generic momenta away from
the conventional high-symmetry path, we additionally performed direct
SOC-included DFT calculations along four short local momentum paths
centered at representative Wannier-identified Weyl-node coordinates
near the \(\Gamma\text{--}Z\) direction. For each node
\(\mathbf{k}_{W_i}=(k_{1,i},k_{2,i},k_{3,i})\), the local path was
chosen along the first reciprocal-lattice direction as
\begin{equation}
(k_{1,i}-0.01,k_{2,i},k_{3,i})
\longrightarrow
(k_{1,i}+0.01,k_{2,i},k_{3,i}),
\label{eq:S_local_weyl_path}
\end{equation}
using 41 equally spaced \(k\) points, such that the center of each
path coincides with the corresponding Wannier-identified node. The
four representative nodes include two nodes with chiral charge
\(-1\), denoted W1 and W2, and two nodes with chiral charge \(+1\),
denoted W3 and W4.

\begin{figure}[htbp]
\centering
\includegraphics[width=0.9\textwidth]{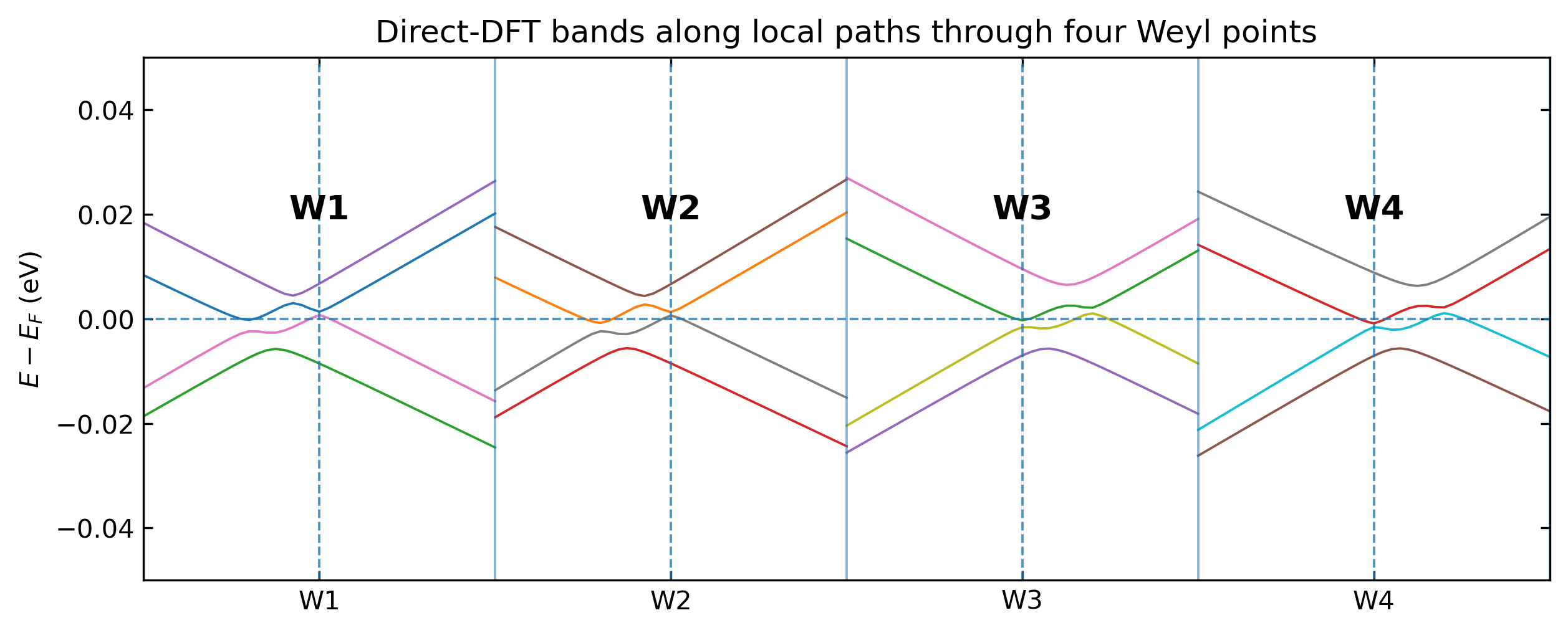}
\caption{
{\bf Direct DFT validation of representative generic-momentum Weyl
crossings.}
SOC-included DFT bands along four short local momentum paths centered
at representative Wannier-identified Weyl nodes in the
\(A(18)\)-distorted structure at
\(Q=2.5~\mathrm{\AA}\sqrt{\mathrm{amu}}\). W1 and W2 carry chiral
charge \(-1\), whereas W3 and W4 carry chiral charge \(+1\), as
determined separately from the Berry-flux analysis. The vertical
dashed lines mark the Wannier-identified node coordinates at the
center of each local path. The direct DFT gaps at these coordinates
are approximately \(0.65\), \(0.65\), \(1.32\), and \(0.70\)~meV for
W1--W4, respectively. The near-degenerate crossings reproduced by
direct DFT provide direct first-principles support that the
generic-momentum crossings are not produced solely by the Wannier
interpolation. Their Weyl character is established independently by
their nonzero integer chiral charges.
}
\label{fig:S_local_weyl_dft}
\end{figure}

As shown in Figure~\ref{fig:S_local_weyl_dft}, the direct DFT
calculations reproduce near-degenerate band crossings close to all
four Wannier-identified node positions. The direct DFT gaps evaluated
at the Wannier node coordinates are approximately \(0.65\), \(0.65\),
\(1.32\), and \(0.70\)~meV for W1--W4, respectively. For W3, the
minimum direct gap along the sampled local path is approximately
\(1.08\)~meV and occurs at a nearby \(k\) point slightly displaced
from the Wannier-predicted coordinate. This small displacement
reflects the finite interpolation uncertainty and the fact that the
direct DFT validation samples a one-dimensional cut through a
three-dimensional band crossing. Nevertheless, the direct DFT
calculations confirm the presence of near-degenerate band crossings
close to the Wannier-identified generic momenta and provide direct
first-principles support that these features are not produced solely
by the Wannier interpolation.

Candidate band-touching points were located by scanning the full
primitive reciprocal cell,
\begin{equation}
(-0.5,-0.5,-0.5)
+u(1,0,0)+v(0,1,0)+w(0,0,1),
\label{eq:S_weyl_search_region}
\end{equation}
with \(0\le u,v,w\le1\), using a
\(35\times35\times35\) initial search grid. A coarser
\(24\times24\times24\) grid recovered the same Weyl-node
configuration within the numerical tolerance, confirming that the
search is converged with respect to the initial grid density. The
chiral charge of each candidate crossing was subsequently evaluated
from the Berry flux through a closed surface enclosing the node.

Candidate crossings identified from the full-Brillouin-zone search
were classified as Weyl nodes only after their nonzero integer chiral
charges were confirmed by the Berry-flux calculations. Combined with
the agreement between the direct DFT and Wannier-interpolated bands
and the direct DFT validation near representative generic-momentum
nodes, this analysis establishes the Weyl-semimetallic band topology
of the \(A(18)\)-distorted structure. The minimum direct gap along
\(\Gamma\text{--}Z\) is used only as a diagnostic of CDW-gap
suppression and is not employed as the sole criterion for Weyl-node
formation.

\begin{table}[htbp]
\caption{
Summary of the Wannierization, interpolation-validation, and
Weyl-node-search settings.
}
\label{tab:S_wannier_settings}
\centering
\begin{ruledtabular}
\begin{tabular}{l c}
Parameter & Value \\
\hline
Initial projections & Ta \(d\), Se \(p\), I \(p\) \\
\texttt{spinors} & \texttt{true} \\
\texttt{num\_bands} & 180 \\
\texttt{num\_wann} & 148 \\
Outer disentanglement window & \(-14\) to \(10\) eV \\
Frozen window & \(0\) to \(10\) eV \\
\texttt{dis\_num\_iter} & 8000 \\
\texttt{dis\_mix\_ratio} & 0.5 \\
\texttt{dis\_conv\_tol} & \(10^{-8}\) \\
\texttt{num\_iter} & 200 \\
DFT--Wannier validation structure
& \(A(18)\), \(Q=2.5~\mathrm{\AA}\sqrt{\mathrm{amu}}\) \\
DFT--Wannier validation path
& Complete high-symmetry path \\
Local direct-DFT validation
& Four paths, 41 \(k\) points per path \\
\texttt{SOC} in \textsc{WannierTools} & 1 \\
\texttt{NumOccupied} in \textsc{WannierTools} & 98 \\
Search region & Full primitive reciprocal cell \\
Weyl-node search grid & \(35\times35\times35\) \\
Checked search grid & \(24\times24\times24\) \\
\texttt{Gap\_threshold} & \(5.0\times10^{-8}\) \\
Chiral charge calculation
& \texttt{WeylChirality\_calc = T} \\
\end{tabular}
\end{ruledtabular}
\end{table}

\vspace{2em}
\hrule
\vspace{1.5em}

% =====================================================================
% ---------------------------------------------------------------------
% III. Coherent-phonon energy and pump fluence
% ---------------------------------------------------------------------
% =====================================================================

\section{Estimate of Coherent-Phonon Energy and Pump Fluence}
\label{sec:fluence}

To connect the frozen-phonon amplitudes used in the main text with experimentally relevant pump scales, we estimate the harmonic energy stored in the coherent phonon mode. For a mass-weighted normal coordinate \(Q\), expressed in units of \(\mathrm{\AA}\sqrt{\mathrm{amu}}\), the harmonic mode energy is
\begin{equation}
E_\nu(Q)
=
\frac{1}{2}\Omega_\nu^2 Q^2
=
\frac{1}{2}(2\pi f_\nu)^2 Q^2 ,
\label{eq:S_mode_energy}
\end{equation}
where \(f_\nu\) is the phonon frequency. In practical units, this expression becomes
\begin{equation}
E_\nu(\mathrm{eV})
\simeq
2.05\times10^{-3}
f_\nu^2(\mathrm{THz})
Q^2(\mathrm{\AA}^2\mathrm{amu}) .
\label{eq:S_mode_energy_units}
\end{equation}

For the CDW amplitude mode A(18), \(f_\nu=2.52\)~THz. The relevant amplitudes discussed in the main text, \(Q=2.0\)--\(2.5~\mathrm{\AA}\sqrt{\mathrm{amu}}\), therefore correspond to coherent-mode energies of approximately \(0.052\)--\(0.081\)~eV per primitive CDW cell. The corresponding maximum Ta displacements are \(0.078\)--\(0.098\)~\AA.

As a simple estimate of the associated heating scale, we convert this coherent-mode energy into an effective lattice temperature increase by assuming complete thermalization over the \(3N\) lattice degrees of freedom of the 22-atom primitive CDW cell:
\begin{equation}
\Delta T_{\mathrm{eff}}
\simeq
\frac{E_\nu}{3Nk_B},
\label{eq:S_delta_T}
\end{equation}
where \(N=22\). This gives \(\Delta T_{\mathrm{eff}}\approx 9\)--\(14\)~K for \(Q=2.0\)--\(2.5~\mathrm{\AA}\sqrt{\mathrm{amu}}\). This temperature scale should be interpreted as an energy-equivalent estimate associated with the target coherent mode rather than as a prediction of the actual pump-induced sample temperature, because only a fraction of the absorbed optical energy is converted into the selected coherent phonon.

We further estimate the mode-energy-equivalent absorbed fluence, which represents a lower-bound absorbed-energy scale under the assumption that the deposited energy is converted entirely into the target coherent phonon. For an illuminated area \(A\) and an effective optical penetration depth \(l_{\mathrm{abs}}\), over which the absorbed energy is deposited, the excited volume is approximated as \(A l_{\mathrm{abs}}\). This volume contains \(A l_{\mathrm{abs}}/V_{\mathrm{cell}}\) primitive CDW cells. Multiplying this number of cells by the coherent-mode energy \(E_\nu\) per primitive cell and dividing by the illuminated area gives
\begin{equation}
F_{\mathrm{abs}}
\sim
\frac{1}{A}
\left(
\frac{A l_{\mathrm{abs}}}{V_{\mathrm{cell}}}
\right)
E_\nu
=
\frac{E_\nu}{V_{\mathrm{cell}}}
l_{\mathrm{abs}},
\label{eq:S_abs_fluence}
\end{equation}
where \(V_{\mathrm{cell}}\) is the primitive CDW cell volume. Using \(V_{\mathrm{cell}}\simeq 6.2\times10^{-28}~\mathrm{m^3}\) and a representative penetration depth \(l_{\mathrm{abs}}=0.1\)--\(1~\mu\mathrm{m}\), we obtain the order-of-magnitude estimates listed in Table~\ref{tab:S_A18_fluence}. The actual incident fluence can be higher than this lower-bound absorbed-fluence scale, depending on reflectivity, absorption coefficient, damping, pulse duration, Raman susceptibility, and coherent-phonon generation efficiency.

\begin{table}[htbp]
\caption{Order-of-magnitude experimental scales for the A(18) coherent phonon amplitude. The mode energy is estimated from \(E_\nu=\frac{1}{2}(2\pi f_\nu)^2Q^2\) with \(f_\nu=2.52\)~THz. The effective temperature increase assumes complete thermalization over the \(3N\) lattice degrees of freedom of the 22-atom primitive CDW cell. The absorbed fluence is a mode-energy-equivalent lower-bound estimate obtained using a representative optical penetration depth of \(0.1\)--\(1~\mu\mathrm{m}\).}
\label{tab:S_A18_fluence}
\centering
\begin{ruledtabular}
\begin{tabular}{c c c c c}
\(Q\) & \(D_{\max}\) & \(E_\nu\) & \(\Delta T_{\mathrm{eff}}\) & \(F_{\mathrm{abs}}\) \\
\((\mathrm{\AA}\sqrt{\mathrm{amu}})\) & \((\mathrm{\AA})\) & (eV/cell) & (K) & \((\mathrm{mJ/cm^2})\) \\
\hline
2.0 & 0.078 & 0.052 & 9  & 0.13--1.3 \\
2.5 & 0.098 & 0.081 & 14 & 0.21--2.1 \\
\end{tabular}
\end{ruledtabular}
\end{table}
These excitation scales are consistent with experimentally accessible ultrafast excitation conditions demonstrated in recent optical-pump experiments~\cite{cheng2024chirality} and strong-field THz-pump experiments~\cite{song2023ultrafast}.
For completeness, we also estimate the corresponding peak absorbed
intensity for a Gaussian pump pulse. Assuming the absorbed intensity
has the temporal profile
\begin{equation}
I_{\mathrm{abs}}(t)
=
I_{\mathrm{abs}}^{\mathrm{peak}}
\exp\left[
-\frac{4\ln 2\,t^2}{\tau_{\mathrm{FWHM}}^2}
\right],
\end{equation}
where \(\tau_{\mathrm{FWHM}}\) is the temporal full width at half
maximum of the intensity profile. This parametrization satisfies
\(I_{\mathrm{abs}}(\pm\tau_{\mathrm{FWHM}}/2)
=I_{\mathrm{abs}}^{\mathrm{peak}}/2\). The absorbed fluence is
\begin{equation}
F_{\mathrm{abs}}
=
\int_{-\infty}^{+\infty}I_{\mathrm{abs}}(t)\,dt
=
I_{\mathrm{abs}}^{\mathrm{peak}}
\tau_{\mathrm{FWHM}}
\sqrt{\frac{\pi}{4\ln 2}}.
\end{equation}
Therefore, the peak absorbed intensity is
\begin{equation}
I_{\mathrm{abs}}^{\mathrm{peak}}
=
\sqrt{\frac{4\ln 2}{\pi}}
\frac{F_{\mathrm{abs}}}{\tau_{\mathrm{FWHM}}}.
\label{eq:S_peak_intensity}
\end{equation}
For an illustrative Gaussian pulse with
\(\tau_{\mathrm{FWHM}}=100~\mathrm{fs}\), the lower-bound
absorbed-fluence range in Table~\ref{tab:S_A18_fluence} corresponds to peak absorbed
intensities of order \(1\)--\(20~\mathrm{GW/cm^2}\).
The actual incident peak intensity may be larger depending on the optical absorption, Raman susceptibility, and coherent-phonon generation efficiency.

These estimates indicate that the \(A(18)\) amplitude needed for strong CDW-gap suppression corresponds to a coherent-mode energy of approximately \(0.05\)--\(0.08\)~eV per primitive CDW cell and to a mode-energy-equivalent lower-bound absorbed fluence on the order of \(10^{-1}\)--a few \(\mathrm{mJ/cm^2}\), depending on the penetration depth. In contrast, the higher-frequency Se-dominated Raman modes require larger normal-mode amplitudes and larger maximum atomic displacements, leading to substantially larger coherent-mode energies. This supports the conclusion that the low-frequency A(18) CDW amplitude mode is the most experimentally feasible Raman pathway for driving the system toward the Weyl-semimetal regime.

\vspace{2em}
\hrule
\vspace{1.5em}

% =====================================================================
% ---------------------------------------------------------------------
% IV. Mode effective charge and THz field strength
% ---------------------------------------------------------------------
% =====================================================================

\section{Mode Effective Charge and THz Field Strength for Driving the $B_3(7)$ Mode}
\label{sec:thz}

To assess whether the IR-active B$_3$(7) mode can be coherently driven to amplitudes relevant for the nonlinear IR--Raman coupling, we estimate its mode effective charge and the THz field scale required to reach a target peak amplitude \(Q_0\) of \(Q_{\mathrm{IR}}(t)\).

The mode effective charge of an IR phonon is obtained by projecting the Born effective charge (BEC) tensors \(Z^{\ast}_{\kappa,\alpha\beta}\) onto the phonon eigenvector. Using the mass-weighted eigenvector \(e_{\kappa\beta}\), normalized according to \(\sum_{\kappa\beta}|e_{\kappa\beta}|^2=1\), obtained from our phonon calculation, the physical displacement pattern per unit normal coordinate is \(u_{\kappa\beta}=e_{\kappa\beta}/\sqrt{M_\kappa}\), and the mode effective charge along Cartesian direction \(\alpha\) is
\begin{equation}
Z^{\ast}_{\mathrm{mode},\alpha}
=
\sum_{\kappa}\sum_{\beta}
Z^{\ast}_{\kappa,\alpha\beta}\,
\frac{e_{\kappa\beta}}{\sqrt{M_\kappa}},
\label{eq:S_zmode}
\end{equation}
where \(\kappa\) labels atoms and \(M_\kappa\) is the atomic mass. The BEC tensors were computed using density-functional perturbation theory for the relaxed \(F222\) CDW structure at the equilibrium geometry without an imposed phonon distortion.

Direct evaluation of Eq.~(\ref{eq:S_zmode}) using the complete atom-resolved BEC tensors and the normalized \(B_3(7)\) eigenvector gives
\begin{equation}
\left(
Z_{\mathrm{mode},x}^{\ast},
Z_{\mathrm{mode},y}^{\ast},
Z_{\mathrm{mode},z}^{\ast}
\right)
\simeq
\left(
0,
0,
1.806325
\right)
~e/\sqrt{\mathrm{amu}},
\label{eq:S_zmode_vector}
\end{equation}
where the \(x\)- and \(y\)-components are below \(1.5\times10^{-6}~e/\sqrt{\mathrm{amu}}\). The mode effective charge is therefore directed almost entirely along the chain direction, with \(z\parallel c\). Because the overall phase of a phonon eigenvector is arbitrary, reversing the eigenvector changes the sign of \(Z_{\mathrm{mode},z}^{\ast}\) but not its magnitude. For the field-amplitude estimate below, we therefore define
\begin{equation}
Z_{\mathrm{mode}}^{\ast}
\equiv
\left|Z_{\mathrm{mode},z}^{\ast}\right|
\approx
1.8~e/\sqrt{\mathrm{amu}}.
\label{eq:S_zmode_value}
\end{equation}
An element-resolved decomposition shows that the mode effective charge is dominated by the Ta contribution, while the Se and I contributions provide smaller negative corrections. The value in Eq.~(\ref{eq:S_zmode_value}) is obtained from the complete projection of the atom-resolved BEC tensors onto the normalized \(B_3(7)\) eigenvector, including all Cartesian displacement components. The element-resolved contributions to \(Z_{\mathrm{mode},z}^{\ast}\) are summarized in Table~\ref{Table:S_mode_charge}.

\begin{table}[htbp]
\caption{Element-resolved contributions to the dominant \(z\)-component of the \(B_3(7)\) mode effective charge. Each value is summed over all atoms of the corresponding element and over all Cartesian displacement components in Eq.~(\ref{eq:S_zmode}).}
\label{Table:S_mode_charge}
\centering
\begin{tabular}{cc}
\hline\hline
Element & \(\Delta Z_{\mathrm{mode},z}^{\ast}\) \((e/\sqrt{\mathrm{amu}})\) \\
\hline
Ta & \(+1.875332\) \\
Se & \(-0.052978\) \\
I & \(-0.016029\) \\
\hline
Total & \(+1.806325\) \\
\hline\hline
\end{tabular}
\end{table}

To clarify the dominant Ta contribution, Table~\ref{Table:S_Ta_mode_charge} lists the longitudinal contributions of the four Ta atoms. For these Ta ions, the off-diagonal contributions to \(Z_{\mathrm{mode},z}^{\ast}\) are numerically negligible. For the two dominant Ta contributions, the signs of both \(Z_{\kappa,zz}^{\ast}\) and \(e_{\kappa z}\) reverse, so their longitudinal contributions have the same sign and add constructively rather than cancel.

\begin{table}[htbp]
\caption{Longitudinal contributions of the four Ta atoms to the \(z\)-component of the \(B_3(7)\) mode effective charge. The ion indices follow the atomic ordering used in the 22-atom primitive-cell calculation and are not intended to denote the site labels used in the structural schematic. The last column is evaluated as \(Z_{\kappa,zz}^{\ast}e_{\kappa z}/\sqrt{M_{\mathrm{Ta}}}\).}
\label{Table:S_Ta_mode_charge}
\centering
\begin{tabular}{cccc}
\hline\hline
Ion & \(Z_{\kappa,zz}^{\ast}\) (\(e\)) & \(e_{\kappa z}\) & \(\Delta Z_{\kappa,z}^{\ast}\) \((e/\sqrt{\mathrm{amu}})\) \\
\hline
Ta (ion 1) & \(-18.50992\) & \(-0.506929\) & \(+0.697549\) \\
Ta (ion 2) & \(+25.34905\) & \(+0.630807\) & \(+1.188725\) \\
Ta (ion 3) & \(+0.67490\) & \(-0.109060\) & \(-0.005472\) \\
Ta (ion 4) & \(+0.67471\) & \(-0.109060\) & \(-0.005470\) \\
\hline
Ta total & & & \(+1.875332\) \\
\hline\hline
\end{tabular}
\end{table}
In the mass-weighted normal coordinate, the kinetic and harmonic potential energies are \(\frac{1}{2}\dot Q_{\mathrm{IR}}^2\) and \(\frac{1}{2}\omega_{\mathrm{IR}}^2Q_{\mathrm{IR}}^2\), respectively. The linear coupling to the THz field is \(-Z_{\mathrm{mode}}^\ast Q_{\mathrm{IR}}E(t)\). The coordinate-dependent potential energy is therefore
\begin{equation}
V(Q_{\mathrm{IR}},t)
=
\frac{1}{2}\omega_{\mathrm{IR}}^2Q_{\mathrm{IR}}^2
-
Z_{\mathrm{mode}}^\ast Q_{\mathrm{IR}}E(t).
\label{eq:S_ir_potential}
\end{equation}
In this mass-weighted normalization, the normal coordinate has unit effective mass, so the force balance is
\begin{equation}
\ddot Q_{\mathrm{IR}}
=
-\frac{\partial V}{\partial Q_{\mathrm{IR}}}
-
2\Gamma\dot Q_{\mathrm{IR}}
=
-\omega_{\mathrm{IR}}^2Q_{\mathrm{IR}}
+
Z_{\mathrm{mode}}^\ast E(t)
-
2\Gamma\dot Q_{\mathrm{IR}}.
\label{eq:S_ir_force_balance}
\end{equation}
Taking \(E(t)=E_0\cos(\omega_{\mathrm{IR}}t)\) and rearranging the damping and restoring-force terms to the left-hand side gives the driven equation of motion
\begin{equation}
\ddot{Q}_{\mathrm{IR}}
+
2\Gamma\dot{Q}_{\mathrm{IR}}
+
\omega_{\mathrm{IR}}^2Q_{\mathrm{IR}}
=
Z_{\mathrm{mode}}^\ast E_0
\cos(\omega_{\mathrm{IR}}t),
\label{eq:S_driven_eom}
\end{equation}
where \(\Gamma\) is the amplitude-decay rate in angular-frequency units. After the transient response has decayed, driving at \(\omega=\omega_{\mathrm{IR}}\) produces a steady-state phonon response that lags the driving field by \(\pi/2\) and can be written as \(Q_{\mathrm{IR}}(t)=Q_0\sin(\omega_{\mathrm{IR}}t)\). Substitution into Eq.~(\ref{eq:S_driven_eom}) gives \(2\Gamma\omega_{\mathrm{IR}}Q_0=Z_{\mathrm{mode}}^\ast E_0\). Therefore, the steady-state phonon amplitude is
\begin{equation}
Q_0
=
\frac{Z_{\mathrm{mode}}^\ast E_0}
{2\omega_{\mathrm{IR}}\Gamma}.
\label{eq:S_driven_amp}
\end{equation}
Here, \(\omega_{\mathrm{IR}}=2\pi f_{\mathrm{IR}}\) and \(f_{\mathrm{IR}}=1.15\)~THz. Solving for the required field gives
\begin{equation}
E_0
\approx
\frac{2\omega_{\mathrm{IR}}\Gamma Q_0}
{Z^{\ast}_{\mathrm{mode}}}.
\label{eq:S_required_field}
\end{equation}

To connect the field estimate directly with the nonlinear-response results presented in Figure~4 of the main text and with the rectification analysis of Sec.~\ref{sec:rectified}, we adopt the same representative peak amplitude of the \(B_3(7)\) coordinate,
\begin{equation}
Q_0
=
2.0~\mathrm{\AA}\sqrt{\mathrm{amu}},
\label{eq:S_b3_benchmark_amplitude}
\end{equation}
where \(Q_0\) denotes the peak amplitude of \(Q_{\mathrm{IR}}(t)\). This value is used as a common benchmark rather than as a threshold amplitude.

To relate the damping parameter in Eq.~(\ref{eq:S_driven_eom}) to a conventional spectral linewidth, we denote by \(\Delta f\) the full width at half maximum of the squared-amplitude response. For a general driving frequency \(\omega\), the steady-state squared amplitude is proportional to
\begin{equation}
\left[
\left(\omega_{\mathrm{IR}}^2-\omega^2\right)^2
+
4\Gamma^2\omega^2
\right]^{-1}.
\label{eq:S_squared_amplitude_response}
\end{equation}
Near resonance,
\begin{equation}
\left(\omega_{\mathrm{IR}}^2-\omega^2\right)^2
\simeq
4\omega_{\mathrm{IR}}^2
\left(\omega-\omega_{\mathrm{IR}}\right)^2,
\qquad
4\Gamma^2\omega^2
\simeq
4\Gamma^2\omega_{\mathrm{IR}}^2.
\label{eq:S_near_resonance_approx}
\end{equation}
Hence, apart from the frequency-independent prefactor
\(1/(4\omega_{\mathrm{IR}}^2)\), the squared-amplitude response reduces
to the Lorentzian form
\begin{equation}
\left[
\left(\omega-\omega_{\mathrm{IR}}\right)^2+\Gamma^2
\right]^{-1}.
\label{eq:S_lorentzian_response}
\end{equation}
The half-maximum condition for this Lorentzian gives
\(\left|\omega-\omega_{\mathrm{IR}}\right|=\Gamma\), so that the two
half-maximum frequencies are
\(\omega_{\mathrm{IR}}-\Gamma\) and
\(\omega_{\mathrm{IR}}+\Gamma\). The angular-frequency full width at
half maximum is therefore
\(\Delta\omega=2\Gamma\). Since
\(\Delta\omega=2\pi\Delta f\), we obtain
\begin{equation}
\Gamma=\pi\Delta f,
\qquad
\mathcal{Q}_{\mathrm{ph}}
=
\frac{\omega_{\mathrm{IR}}}{2\Gamma}
=
\frac{f_{\mathrm{IR}}}{\Delta f},
\label{eq:S_linewidth_relation}
\end{equation}
where \(\mathcal{Q}_{\mathrm{ph}}\) is the phonon quality factor, defined as the ratio of the resonance frequency to the full width at half maximum of the squared-amplitude response.

Because the linewidth of the \(B_3(7)\) mode has not been determined in the present harmonic phonon calculation, we consider an illustrative phonon quality-factor range \(\mathcal{Q}_{\mathrm{ph}}=10\)--\(40\). Using \(f_{\mathrm{IR}}=1.15\)~THz, this range corresponds to
\begin{equation}
\Delta f
=
\frac{f_{\mathrm{IR}}}{\mathcal{Q}_{\mathrm{ph}}}
\simeq
0.029\text{--}0.115~\mathrm{THz}.
\label{eq:S_linewidth_range}
\end{equation}

For transparency in the unit conversion, we use
\begin{equation}
1~
\frac{\mathrm{amu}\,\mathrm{\AA}}
{e\,\mathrm{ps}^{2}}
=
10.36~\mathrm{kV/cm}.
\label{eq:S_field_unit_conversion}
\end{equation}
Substituting \(\omega_{\mathrm{IR}}=2\pi f_{\mathrm{IR}}\) and \(\Gamma=\pi\Delta f\) into Eq.~(\ref{eq:S_required_field}) gives
\begin{equation}
E_0(\mathrm{kV/cm})
\approx
4\pi^2(10.36)
\frac{
f_{\mathrm{IR}}(\mathrm{THz})\,
\Delta f(\mathrm{THz})\,
Q_0(\mathrm{\AA}\sqrt{\mathrm{amu}})
}{
Z_{\mathrm{mode}}^\ast(e/\sqrt{\mathrm{amu}})
}.
\label{eq:S_required_field_practical}
\end{equation}
Using \(f_{\mathrm{IR}}=1.15\)~THz, \(\Delta f=0.029\)--\(0.115\)~THz, \(Q_0=2.0~\mathrm{\AA}\sqrt{\mathrm{amu}}\), and \(Z_{\mathrm{mode}}^\ast=1.8~e/\sqrt{\mathrm{amu}}\), we obtain
\begin{equation}
E_0
\simeq
15\text{--}60~\mathrm{kV/cm}.
\label{eq:S_field_value}
\end{equation}
Thus, the required steady-state resonant field is of order tens of \(\mathrm{kV/cm}\).

This field scale should be regarded as a steady-state resonant order-of-magnitude estimate rather than as a quantitative prediction for a specific experimental pulse. For a finite-duration THz pulse, the required peak field may be larger and depends on the pulse duration, spectral bandwidth, detuning from the phonon resonance, damping rate, and temporal field profile. The estimated steady-state field scale is below experimentally demonstrated strong-field THz peak fields, which can exceed \(0.1~\mathrm{MV/cm}\) and reach the \(\mathrm{MV/cm}\) regime~\cite{novelli2020towards,hirori2011single}, suggesting that substantial coherent excitation of B$_3$(7) may be experimentally accessible under favorable resonant conditions.

We emphasize that this estimate addresses only the direct drivability of the B$_3$(7) coordinate and does not imply that the resulting nonlinear coupling is sufficient to reach the CDW-gap-suppression regime. As shown quantitatively in Sec.~\ref{sec:rectified}, at this benchmark amplitude the rectified \(A(18)\) displacement is directed toward the CDW-suppressing branch but remains far below the magnitude required for strong gap suppression.

\vspace{2em}
\hrule
\vspace{1.5em}

% =====================================================================
% ---------------------------------------------------------------------
% V. Rectified A(18) Displacement Induced by the B$_3$(7)
% ---------------------------------------------------------------------
% =====================================================================

\section{Estimate of the Rectified $A(18)$ Displacement Induced by the $B_3(7)$ Mode}
\label{sec:rectified}

To assess quantitatively whether coherent excitation of the IR-active B$_3$(7) mode can produce a unidirectional displacement of the Raman-active CDW amplitude mode A(18), we estimate the cycle-averaged rectified A(18) response within a quasi-static approximation. We denote the Raman coordinate by $Q_R\equiv Q_{A(18)}$ and the IR coordinate by $Q_{\mathrm{IR}}\equiv Q_{B_3(7)}$.

\subsection{Normal-mode normalization}

The displacement patterns used to construct the nonlinear potential-energy surfaces were generated from phonon eigenvectors normalized in the 22-atom primitive CDW cell and then mapped onto the two periodically equivalent primitive-cell sectors of the 44-atom conventional cell used for the total-energy calculations. The fitting coordinates retained the 22-atom normal-coordinate convention, while the calculated energy differences were normalized per atom. Since the 44-atom conventional cell contains two equivalent 22-atom primitive cells,
\begin{equation}
\Delta E^{(44)}(Q)=2\Delta E^{(22)}(Q),
\end{equation}
and therefore
\begin{equation}
\frac{\Delta E^{(44)}(Q)}{44}=\frac{\Delta E^{(22)}(Q)}{22}.
\label{eq:S_energy_normalization}
\end{equation}
Thus, all fitted anharmonic coefficients and normal coordinates used below consistently employ the 22-atom primitive-cell convention together with a per-atom energy normalization.

\subsection{Nonlinear potential-energy surface}

Following the coefficient convention of Eq.~(2) in the main text, the symmetry-allowed nonlinear potential per atom is written as
\begin{equation}
\begin{split}
V(Q_{\mathrm{IR}},Q_R)={}&\frac{1}{2}K_{\mathrm{IR}}Q_{\mathrm{IR}}^2+\frac{1}{2}K_RQ_R^2+\frac{c_2}{3}Q_R^3+c_{21}Q_{\mathrm{IR}}^2Q_R+\frac{d_1}{4}Q_{\mathrm{IR}}^4+\frac{d_2}{4}Q_R^4+d_{22}Q_{\mathrm{IR}}^2Q_R^2+\cdots .
\end{split}
\label{eq:S_pes_rectified}
\end{equation}
Here, $K_{\mathrm{IR}}$ and $K_R$ denote the harmonic coefficients expressed using the same per-atom energy normalization and 22-atom normal-coordinate convention as the fitted anharmonic coefficients.

For the B$_3$(7)--A(18) pair, the fitted coefficients are
\begin{equation}
c_2=-0.180,\qquad c_{21}=-0.080,\qquad d_1=-0.002,\qquad d_2=-0.019,\qquad d_{22}=0.009.
\label{eq:S_coeff_b3_a18}
\end{equation}
The coefficients are expressed in
\begin{equation}
\frac{\mathrm{meV\,atom^{-1}}}{\mathrm{amu}^{k/2}\mathrm{\AA}^{k}},
\end{equation}
where $k$ is the total order of the corresponding displacement term. Thus, $c_2$ and $c_{21}$ have $k=3$, whereas $d_1$, $d_2$, and $d_{22}$ have $k=4$. The numerical prefactors $1/3$ and $1/4$ in Eq.~(\ref{eq:S_pes_rectified}) follow exactly the definition used in the main text.

\subsection{Cycle-averaged rectification}

For a coherently driven IR mode, we write
\begin{equation}
Q_{\mathrm{IR}}(t)
=
Q_0\cos(\Omega_{\mathrm{IR}}t),
\label{eq:S_qir_time}
\end{equation}
where \(Q_0\) is the peak IR-mode amplitude. The squared IR-mode coordinate is
\begin{equation}
Q_{\mathrm{IR}}^2(t)
=
Q_0^2\cos^2(\Omega_{\mathrm{IR}}t)
=
\frac{Q_0^2}{2}
\left[
1+\cos(2\Omega_{\mathrm{IR}}t)
\right].
\label{eq:S_qir_squared}
\end{equation}
Averaging over one complete IR oscillation period gives
\begin{equation}
\left\langle
\cos(2\Omega_{\mathrm{IR}}t)
\right\rangle
=0,
\end{equation}
and therefore
\begin{equation}
\left\langle Q_{\mathrm{IR}}^2\right\rangle
=
\frac{Q_0^2}{2}.
\label{eq:S_qir_average}
\end{equation}

To isolate the static component of the IR-induced modification of the
Raman potential, we average the potential over one complete IR
oscillation period while treating \(Q_R\) as a fixed parameter. The
cycle average consequently acts only on the explicitly time-dependent
factors involving \(Q_{\mathrm{IR}}(t)\). In particular,
\begin{equation}
\left\langle
Q_{\mathrm{IR}}^2(t)Q_R
\right\rangle
=
\left\langle
Q_{\mathrm{IR}}^2
\right\rangle Q_R,
\end{equation}
and similarly
\begin{equation}
\left\langle
Q_{\mathrm{IR}}^2(t)Q_R^2
\right\rangle
=
\left\langle
Q_{\mathrm{IR}}^2
\right\rangle Q_R^2.
\end{equation}
Thus, within this cycle-averaged construction,
\begin{equation}
\frac{\partial\langle V\rangle}{\partial Q_R}
=
\left\langle
\frac{\partial V}{\partial Q_R}
\right\rangle.
\end{equation}

Here \(Q_R\) denotes the general normal-mode coordinate of the
\(A(18)\) Raman mode. Under coherent excitation of the IR mode, the
minimum of the cycle-averaged potential may shift away from its
undriven position at \(Q_R=0\). We denote this shift by
\(\overline{Q_R^{\mathrm{ind}}}\), where the overline indicates the
cycle-averaged component and the superscript ``ind'' indicates that
the displacement is induced by the driven IR mode. Since the
undriven minimum is located at \(Q_R=0\),
\(\overline{Q_R^{\mathrm{ind}}}\) is equal to the value of \(Q_R\) at
the minimum of the cycle-averaged driven potential.

Within the quasi-static approximation, this displaced equilibrium
position is obtained by minimizing the cycle-averaged potential:
\begin{equation}
\begin{split}
0
={}&
\left.
\frac{\partial\langle V\rangle}{\partial Q_R}
\right|_{Q_R=\overline{Q_R^{\mathrm{ind}}}}
\\
={}&
K_R\overline{Q_R^{\mathrm{ind}}}
+c_2\left(\overline{Q_R^{\mathrm{ind}}}\right)^2
+d_2\left(\overline{Q_R^{\mathrm{ind}}}\right)^3
+c_{21}\left\langle Q_{\mathrm{IR}}^2\right\rangle
+2d_{22}\left\langle Q_{\mathrm{IR}}^2\right\rangle
\overline{Q_R^{\mathrm{ind}}}.
\end{split}
\label{eq:S_minimize_qr}
\end{equation}

In the small-displacement limit, \(\overline{Q_R^{\mathrm{ind}}}\) is assumed to remain sufficiently close to the equilibrium position that the Raman self-anharmonic contributions
\(c_2\left(\overline{Q_R^{\mathrm{ind}}}\right)^2\) and
\(d_2\left(\overline{Q_R^{\mathrm{ind}}}\right)^3\)
can be neglected. Equation~(\ref{eq:S_minimize_qr}) then reduces to
\begin{equation}
0\simeq
\left[
K_R
+2d_{22}\left\langle Q_{\mathrm{IR}}^2\right\rangle
\right]
\overline{Q_R^{\mathrm{ind}}}
+c_{21}\left\langle Q_{\mathrm{IR}}^2\right\rangle.
\label{eq:S_minimize_qr_small}
\end{equation}
Solving this linearized stationarity condition gives
\begin{equation}
\overline{Q_R^{\mathrm{ind}}}
\simeq
-\frac{
c_{21}\left\langle Q_{\mathrm{IR}}^2\right\rangle
}{
K_R
+2d_{22}\left\langle Q_{\mathrm{IR}}^2\right\rangle
}.
\label{eq:S_qr_induced_quartic}
\end{equation}
The approximate sign reflects the neglect of the Raman self-anharmonic terms in Eq.~(\ref{eq:S_minimize_qr}); within the resulting linearized stationarity equation, the expression is exact.

Using
\(\left\langle Q_{\mathrm{IR}}^2\right\rangle=Q_0^2/2\),
Eq.~(\ref{eq:S_qr_induced_quartic}) becomes
\begin{equation}
\overline{Q_R^{\mathrm{ind}}}
\simeq
-\frac{
(c_{21}/2)Q_0^2
}{
K_R+d_{22}Q_0^2
}.
\label{eq:S_qr_induced_peak}
\end{equation}
If the quartic intermode coupling \(d_{22}\) is also neglected, the lowest-order result is
\begin{equation}
\overline{Q_R^{\mathrm{ind}}}
\simeq
-\frac{c_{21}}{2K_R}Q_0^2.
\label{eq:S_qr_induced_lowest}
\end{equation}

The cycle-averaged force acting on the Raman coordinate at a general
position \(Q_R\) is given by the negative gradient of the
cycle-averaged potential:
\begin{equation}
\begin{split}
\overline{F_R}(Q_R)
={}&
-\frac{\partial\langle V\rangle}{\partial Q_R}
\\
={}&
-K_RQ_R
-c_2Q_R^2
-d_2Q_R^3
-c_{21}\left\langle Q_{\mathrm{IR}}^2\right\rangle
-2d_{22}\left\langle Q_{\mathrm{IR}}^2\right\rangle Q_R .
\end{split}
\label{eq:S_force_general}
\end{equation}

At the undriven equilibrium position \(Q_R=0\), all terms in
Eq.~(\ref{eq:S_force_general}) proportional to \(Q_R\), \(Q_R^2\), or
\(Q_R^3\) vanish. The remaining leading cycle-averaged rectified force
is therefore
\begin{equation}
\overline{F_R}(0)
=
-\left.
\frac{\partial\langle V\rangle}{\partial Q_R}
\right|_{Q_R=0}
=
-c_{21}\left\langle Q_{\mathrm{IR}}^2\right\rangle .
\label{eq:S_rectified_force}
\end{equation}

In the coordinate convention adopted in the main text, positive
\(Q_R\) reduces the Ta-chain tetramerization and moves the structure
toward the nearly equal-spacing non-CDW geometry. Since \(c_{21}<0\)
for the B$_3$(7)--A(18) pair and
\(\langle Q_{\mathrm{IR}}^2\rangle>0\),
Eq.~(\ref{eq:S_rectified_force}) gives
\(\overline{F_R}(0)>0\), showing that the leading rectified force
points toward positive \(Q_R\). Consistently, because \(K_R>0\),
Eq.~(\ref{eq:S_qr_induced_lowest}) gives
\(\overline{Q_R^{\mathrm{ind}}}>0\). Thus, both the rectified force
and the resulting lowest-order induced displacement are directed
toward the structural branch that suppresses the CDW distortion and
reduces the CDW gap.

\subsection{Harmonic coefficient in practical units}

For a mass-weighted normal coordinate, the harmonic mode energy is
\begin{equation}
E_R=\frac{1}{2}(2\pi f_R)^2Q_R^2.
\label{eq:S_harmonic_energy}
\end{equation}
Using \(1~\mathrm{amu}\,\mathrm{\AA^2}\,\mathrm{ps^{-2}}=0.103642~\mathrm{meV}\) and \(1~\mathrm{THz}=1~\mathrm{ps^{-1}}\), the total harmonic coefficient for one 22-atom primitive CDW cell is
\begin{equation}
K_R^{(22),\mathrm{cell}}=(2\pi)^2(0.103642)f_R^2\simeq4.09f_R^2~\frac{\mathrm{meV}}{\mathrm{amu}\,\mathrm{\AA}^2}.
\label{eq:S_kr_conversion}
\end{equation}
For the A(18) frequency $f_R=2.52$~THz,
\begin{equation}
K_R^{(22),\mathrm{cell}}\simeq4.09(2.52)^2\simeq26.0~\frac{\mathrm{meV}}{\mathrm{amu}\,\mathrm{\AA}^2}.
\label{eq:S_kr_cell}
\end{equation}

Because the nonlinear potential is expressed in meV per atom and the fitting coordinates use the 22-atom primitive-cell convention, the corresponding per-atom harmonic coefficient is
\begin{equation}
K_R=\frac{K_R^{(22),\mathrm{cell}}}{22}=\frac{26.0}{22}\simeq1.182~\frac{\mathrm{meV\,atom^{-1}}}{\mathrm{amu}\,\mathrm{\AA}^2}.
\label{eq:S_kr_per_atom}
\end{equation}
This value has the same per-atom energy normalization and 22-atom normal-coordinate convention as the fitted anharmonic coefficients in Eq.~(\ref{eq:S_coeff_b3_a18}).

\subsection{Numerical estimate for an experimentally accessible drive}

As discussed in Sec.~\ref{sec:thz}, projecting the Born effective charges onto the B$_3$(7) mode gives a mode effective charge of approximately $1.8~e/\sqrt{\mathrm{amu}}$. Within the same 22-atom coordinate convention used for the nonlinear-potential fit, a representative peak amplitude
\begin{equation}
Q_0=2.0~\mathrm{\AA}\sqrt{\mathrm{amu}}
\label{eq:S_q0_22}
\end{equation}
corresponds to an estimated steady-state resonant THz field scale of
approximately $15$--$60~\mathrm{kV/cm}$.

Using $Q_0=2.0~\mathrm{\AA}\sqrt{\mathrm{amu}}$, the lowest-order expression in Eq.~(\ref{eq:S_qr_induced_lowest}) gives
\begin{equation}
\overline{Q_R^{\mathrm{ind}}}\approx-\frac{-0.080}{2(1.182)}(2.0)^2\simeq+0.135~\mathrm{\AA}\sqrt{\mathrm{amu}}.
\label{eq:S_qr_lowest_numeric}
\end{equation}
Retaining the positive intermode quartic coefficient $d_{22}=0.009$ gives
\begin{equation}
\overline{Q_R^{\mathrm{ind}}}\approx-\frac{(-0.080/2)(2.0)^2}{1.182+0.009(2.0)^2}\simeq+0.131~\mathrm{\AA}\sqrt{\mathrm{amu}}.
\label{eq:S_qr_d22_numeric}
\end{equation}

The Raman self-anharmonic terms can also be retained by solving the complete quasi-static stationarity condition,
\begin{equation}
d_2Q_R^3+c_2Q_R^2+\left(K_R+d_{22}Q_0^2\right)Q_R+\frac{c_{21}}{2}Q_0^2=0.
\label{eq:S_full_static_equation}
\end{equation}
For $Q_0=2.0~\mathrm{\AA}\sqrt{\mathrm{amu}}$, the stable solution continuously connected to the equilibrium minimum at $Q_R=0$ is
\begin{equation}
\overline{Q_R^{\mathrm{ind}}}\simeq+0.134~\mathrm{\AA}\sqrt{\mathrm{amu}}.
\label{eq:S_qr_full_numeric}
\end{equation}
The other real roots of Eq.~(\ref{eq:S_full_static_equation}) are unstable stationary points with negative local curvature and lie far outside the displacement range used to fit the fourth-order potential. They are therefore not physically relevant to the stable branch continuously connected to the equilibrium minimum.

The lowest-order, $d_{22}$-corrected, and complete quasi-static estimates therefore consistently give a positive rectified displacement of order
\begin{equation}
\overline{Q_R^{\mathrm{ind}}}\sim+0.13~\mathrm{\AA}\sqrt{\mathrm{amu}}
\end{equation}
in the 22-atom primitive-cell normalization.

\subsection{Comparison with the A(18) amplitudes required for gap suppression}

In the main text, an A(18) amplitude of
\begin{equation}
Q_R=2.0~\mathrm{\AA}\sqrt{\mathrm{amu}}
\end{equation}
nearly removes the Ta-chain bond disproportionation and already produces Weyl nodes, while the minimum direct $\Gamma$--Z gap reaches its smallest value at
\begin{equation}
Q_c=2.5~\mathrm{\AA}\sqrt{\mathrm{amu}}.
\end{equation}
Both amplitudes use the same 22-atom primitive-cell normalization as the nonlinear-potential fit.

The complete quasi-static rectified displacement, $\overline{Q_R^{\mathrm{ind}}}\simeq+0.134~\mathrm{\AA}\sqrt{\mathrm{amu}}$, is only approximately $5.4$--$6.7\%$ of the positive A(18) amplitudes relevant for strong suppression of the Ta-chain bond disproportionation and the CDW gap. Equivalently, its magnitude is smaller by a factor of approximately fifteen to nineteen.

Importantly, the induced displacement has the favorable sign: B$_3$(7) rectifies A(18) toward positive $Q_R$, which is the direction that reduces the Ta-chain tetramerization and suppresses the CDW gap. Nevertheless, at the representative drive amplitude considered here, the induced displacement remains far below the $Q_R=2.0$--$2.5~\mathrm{\AA}\sqrt{\mathrm{amu}}$ range relevant for strong CDW-gap suppression and Weyl-state restoration. Thus, coherent excitation of B$_3$(7) alone is unlikely to provide a standalone pathway to the Weyl-semimetallic regime, although the rectified force may assist or bias the A(18) response toward the desired structural direction.

\subsection{Curvature modulation and limitations of the quasi-static estimate}

The positive $d_{22}$ coefficient produces a positive contribution to the curvature of the A(18) potential at the undistorted coordinate $Q_R=0$:
\begin{equation}
K_R^{\mathrm{origin}}=K_R+2d_{22}\left\langle Q_{\mathrm{IR}}^2\right\rangle=K_R+d_{22}Q_0^2.
\label{eq:S_kr_origin}
\end{equation}
This term is consistent with the increased curvature around $Q_R=0$ observed in the frozen two-mode potential-energy surface.

However, the curvature governing oscillations about the displaced quasi-static minimum also contains contributions from the Raman self-anharmonic terms:
\begin{equation}
K_R^{\mathrm{local}}=K_R+d_{22}Q_0^2+2c_2\overline{Q_R^{\mathrm{ind}}}+3d_2\left(\overline{Q_R^{\mathrm{ind}}}\right)^2.
\label{eq:S_kr_local}
\end{equation}
Consequently, the sign and magnitude of the actual pump-induced A(18) frequency shift cannot be inferred from $d_{22}$ alone without explicitly accounting for the displaced equilibrium position and the time-dependent lattice dynamics.

The present estimate retains only the cycle-averaged component of $Q_{\mathrm{IR}}^2(t)$. The remaining oscillatory component acts on A(18) at twice the IR frequency. For B$_3$(7),
\begin{equation}
2f_{B_3(7)}=2.30~\mathrm{THz},
\end{equation}
which is relatively close to the A(18) frequency of $2.52$~THz. The oscillatory force may therefore contribute an enhanced transient A(18) response that is not captured by the quasi-static treatment. Accordingly, the quasi-static result should be interpreted specifically as an estimate of the cycle-averaged static displacement, rather than as a prediction of the full time-dependent A(18) response.

A quantitative description of this transient component would require explicit driven lattice-dynamics simulations including a specified THz pulse shape, phonon damping, and a potential-energy surface that remains stable over the full amplitude range sampled by the trajectory. Such a calculation is beyond the scope of the present quasi-static estimate.

We therefore conclude specifically that the cycle-averaged rectification generated by B$_3$(7) is directed toward the CDW-suppressing A(18) coordinate. However, at the representative experimentally accessible drive amplitude, its magnitude remains substantially below the A(18) amplitudes required for strong CDW-gap suppression and Weyl-state restoration. The nonlinear coupling may therefore assist or bias the A(18) response in the favorable direction and may also contribute to transient mode-mixing effects under resonant THz excitation, but B$_3$(7) excitation alone is unlikely to provide a standalone pathway to the Weyl-semimetallic regime.

\begin{figure}[htbp]
\centering
\includegraphics[width=0.60\textwidth]{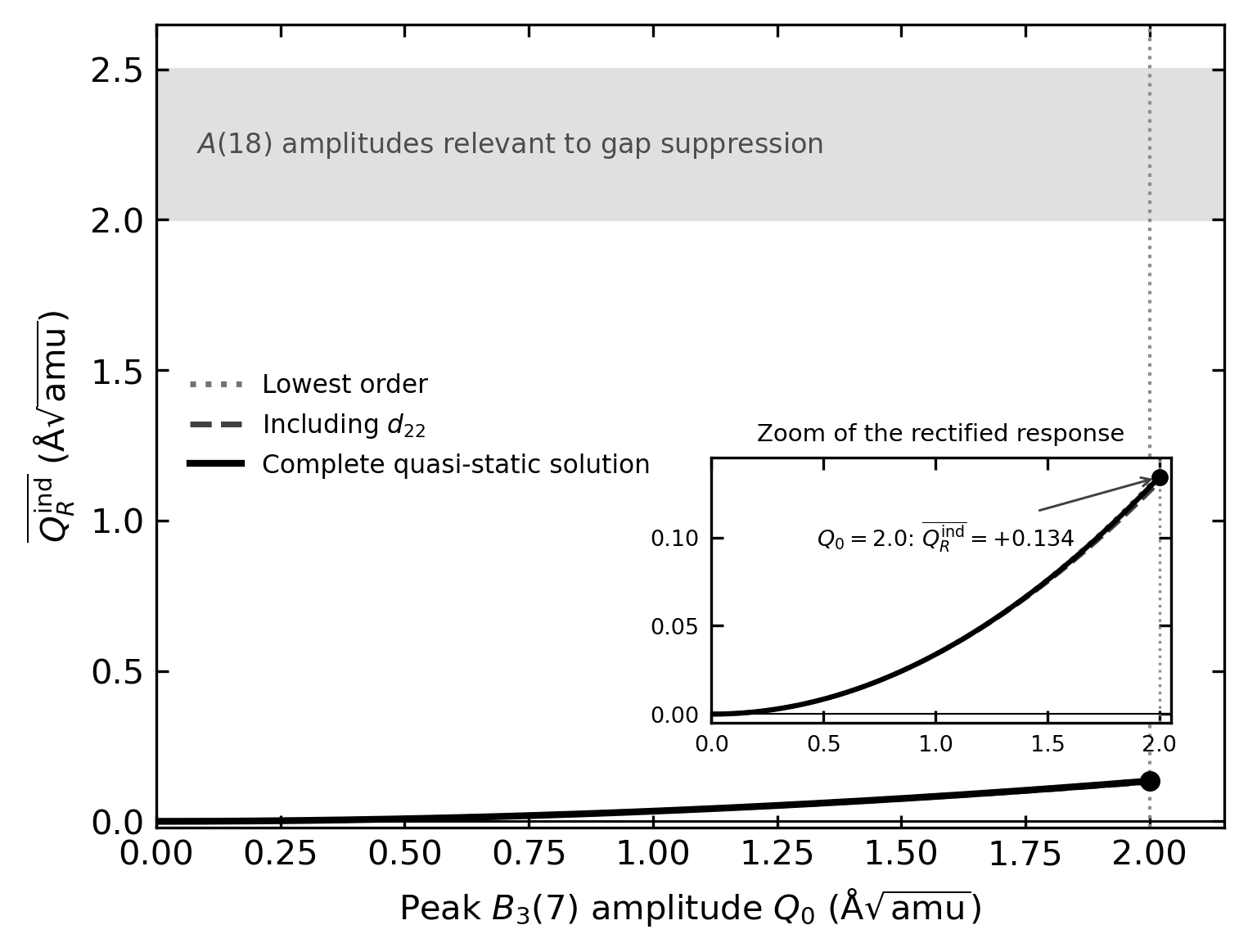}
\caption{Quasi-static rectified A(18) response induced by coherent excitation of the B$_3$(7) mode. The dotted curve shows the lowest-order result from Eq.~(\ref{eq:S_qr_induced_lowest}), the dashed curve includes the quartic intermode coupling $d_{22}$ through Eq.~(\ref{eq:S_qr_induced_peak}), and the solid curve is the stable solution of the complete stationarity condition in Eq.~(\ref{eq:S_full_static_equation}). All coordinates use the 22-atom primitive-cell normalization employed in fitting the nonlinear potential-energy surface. The shaded region marks the positive A(18) amplitudes relevant to CDW-gap suppression: $Q_R=2.0~\mathrm{\AA}\sqrt{\mathrm{amu}}$, where the Ta-chain disproportionation is nearly removed and Weyl nodes are already present, and $Q_c=2.5~\mathrm{\AA}\sqrt{\mathrm{amu}}$, where the minimum direct $\Gamma\text{--}Z$ gap reaches its smallest value. The vertical dotted line at $Q_0=2.0~\mathrm{\AA}\sqrt{\mathrm{amu}}$ corresponds to the estimated steady-state resonant THz field scale of $15$--$60~\mathrm{kV/cm}$ discussed in Sec.~\ref{sec:thz}. At this amplitude, the complete quasi-static solution is $\overline{Q_R^{\mathrm{ind}}}\simeq+0.134~\mathrm{\AA}\sqrt{\mathrm{amu}}$. The inset enlarges the small positive rectified response and shows that the three levels of approximation give closely consistent results. The curves represent the static component obtained from the cycle-averaged quasi-static treatment and do not include the time-dependent response driven by the oscillatory component at \(2\Omega_{\mathrm{IR}}\).}
\label{fig:S_rectified}
\end{figure}

\vspace{2em}
\hrule
\vspace{1.5em}

% =====================================================================
% ---------------------------------------------------------------------
% VI. Phonon Dispersion of A(18) 
% ---------------------------------------------------------------------
% =====================================================================
\section{Full Phonon Dispersion and Dynamical Stability}
\label{sec:dispersion}

To assess the dynamical stability of the relaxed \(F222\) CDW
structure, we calculated the full phonon dispersion using the
finite-displacement method. A \(2\times2\times2\) supercell was
constructed from the 22-atom primitive CDW cell, corresponding to a
total of 176 atoms. The forces for the displaced supercell structures
were evaluated using a \(\Gamma\)-centered \(2\times2\times2\)
\(k\)-point mesh, and the resulting real-space force constants were
used to interpolate the phonon frequencies along the sampled
high-symmetry paths of the primitive Brillouin zone.

\begin{figure}[htbp]
    \centering
    \includegraphics[width=0.92\linewidth]{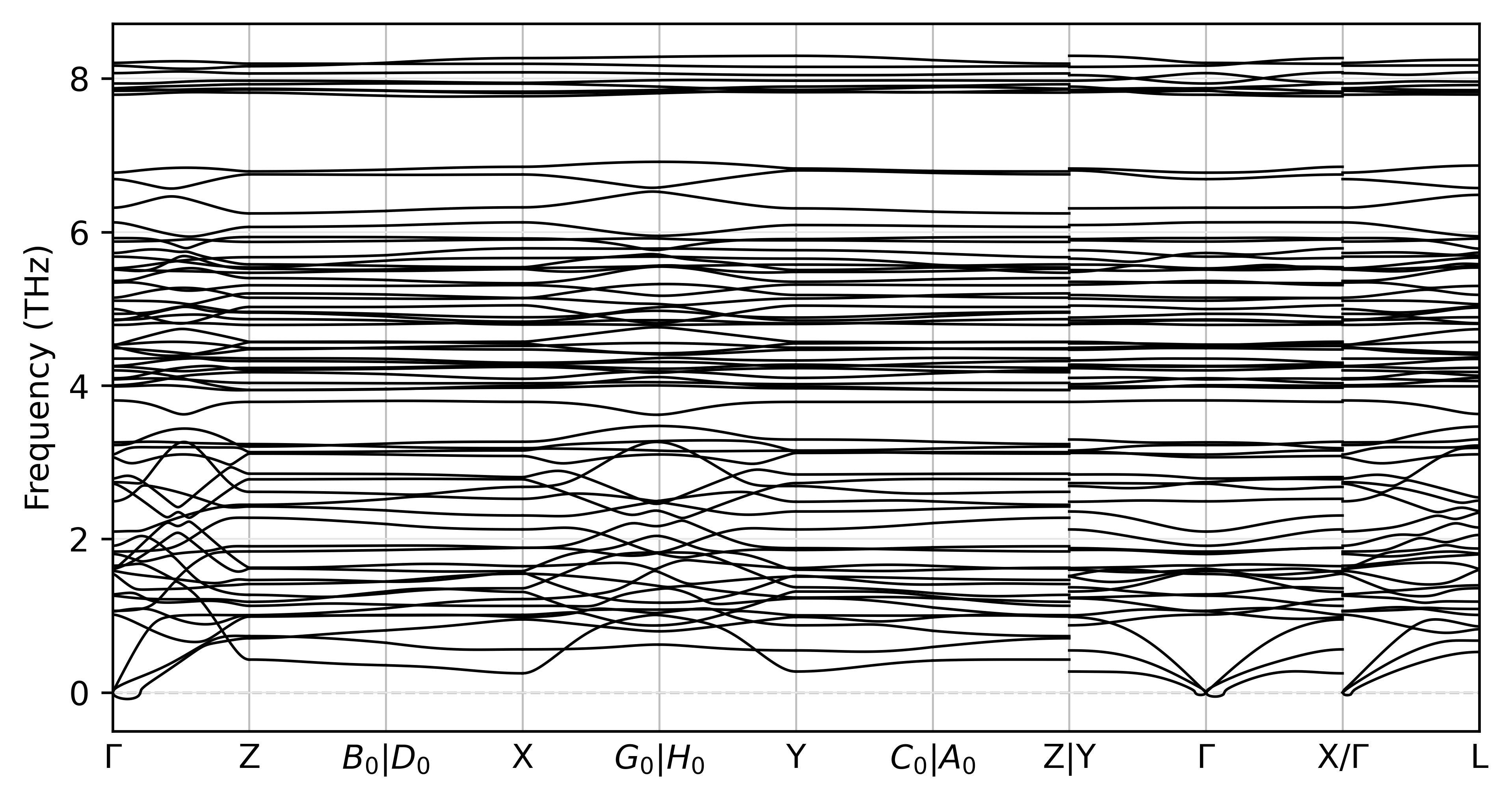}
    \caption{
    Full phonon dispersion of the relaxed \(F222\) CDW structure
    calculated using the finite-displacement method with a
    \(2\times2\times2\) supercell of the 22-atom primitive cell,
    corresponding to 176 atoms. The forces for the displaced
    supercell structures were evaluated using a \(\Gamma\)-centered
    \(2\times2\times2\) \(k\)-point mesh. All optical branches remain
    positive along the sampled high-symmetry paths. Only very small
    negative frequencies below approximately \(0.1\)~THz occur in
    the acoustic branches in the immediate vicinity of \(\Gamma\),
    which are attributed to residual numerical errors in the force
    constants and their interpolation rather than to a genuine
    structural instability.
    }
    \label{fig:S_full_phonon_dispersion}
\end{figure}

The calculated phonon dispersion is shown in
Figure~\ref{fig:S_full_phonon_dispersion}. No physically meaningful
imaginary phonon branches are observed along the sampled
high-symmetry paths. All optical branches remain at positive
frequencies throughout the dispersion. Only very small negative
frequencies, with magnitudes below approximately \(0.1\)~THz, occur
in the acoustic branches in the immediate vicinity of \(\Gamma\).
Because translational invariance requires the three acoustic modes to
approach zero frequency at \(\Gamma\), these shallow negative values
are attributed to residual numerical errors in the finite-displacement
force constants and their interpolation rather than to a genuine
lattice instability. The relaxed \(F222\) CDW structure can therefore
be regarded as dynamically stable within the numerical accuracy and
supercell size employed in the present calculations.

\vspace{2em}
\hrule
\vspace{1.5em}

% =====================================================================
% ---------------------------------------------------------------------
% VII. PBEsol validation
% ---------------------------------------------------------------------
% =====================================================================

\section{PBEsol Validation of Phonon-Driven Weyl-State Restoration}
\label{sec:pbesol}

\begin{figure}[htbp]
\centering
\includegraphics[width=\textwidth]{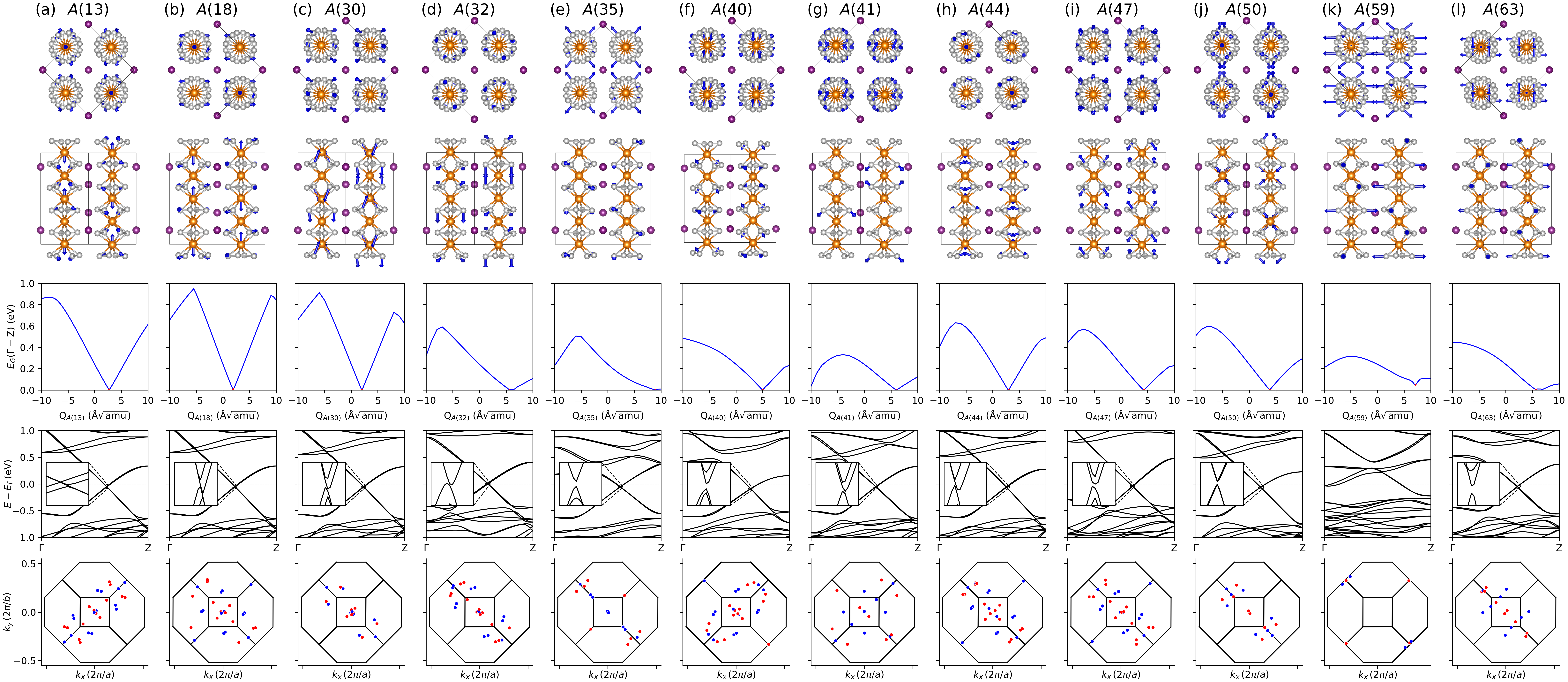}
\caption{
{\bf PBEsol validation of phonon-driven Weyl-state restoration in the $F222$ CDW phase of (TaSe$_4$)$_2$I.}
(a)--(l) Results for twelve symmetry-preserving Raman-active $A$ modes: $A(13)$, $A(18)$, $A(30)$, $A(32)$, $A(35)$, $A(40)$, $A(41)$, $A(44)$, $A(47)$, $A(50)$, $A(59)$, and $A(63)$, respectively. For each mode, the first and second rows show the phonon eigenvectors in the top and side views, respectively; blue arrows indicate the atomic displacement directions. Orange, gray, and purple spheres denote Ta, Se, and I atoms, respectively. The third row shows the minimum direct $\Gamma\text{--}\mathrm{Z}$ gap as a function of the normal-mode coordinate $Q$; the fourth row shows the corresponding band structure at the near-gap-closure displacement; and the bottom row shows the Weyl-node distribution projected onto the $k_x\text{--}k_y$ plane. Red and blue points denote Weyl nodes with positive and negative chirality, respectively.
}
\label{fig:S_pbesol_modes}
\end{figure}

To assess the robustness of the phonon-driven gap suppression against the choice of exchange--correlation functional, we repeated the zone-center phonon calculations and frozen-phonon electronic-structure analysis using the PBEsol functional~\cite{perdew2008restoring}. The PBEsol phonon spectrum retains the same \(F222\)-based symmetry classification, in which Raman-active \(A\) modes correspond to symmetry-preserving lattice distortions. As summarized in Table~\ref{tab:S_pbesol_phonons} and Figure~\ref{fig:S_pbesol_modes}, we identify twelve symmetry-preserving Raman-active \(A\) modes in the PBEsol calculations---\(A(13)\), \(A(18)\), \(A(30)\), \(A(32)\), \(A(35)\), \(A(40)\), \(A(41)\), \(A(44)\), \(A(47)\), \(A(50)\), \(A(59)\), and \(A(63)\)---that strongly suppress the minimum direct gap along the \(\Gamma\text{--}Z\) path and generate Weyl nodes at generic \(k\) points. Although the detailed phonon frequencies, gap-minimizing amplitudes, and mode ordering differ from the PBE results in the main text, the qualitative conclusion remains unchanged: symmetry-preserving Raman distortions can strongly suppress the CDW gap and access a Weyl-semimetallic regime, demonstrating that the principal conclusions of the main text are robust with respect to the choice of exchange--correlation functional.

Table~\ref{tab:S_pbesol_qc} lists the gap-minimizing amplitudes and the corresponding energetic scales for these twelve modes. Eleven of them reduce the $\Gamma\text{--}Z$ direct gap to the meV scale ($E_G^{\min}=0.4$--$3.4$~meV), while $A(59)$ retains a noticeably larger residual gap of $50.5$~meV and is therefore a less effective gap-suppression channel despite also producing Weyl nodes.

The quantitative conclusion drawn in the main text from the PBE results carries over. The two modes carrying CDW-amplitude character, $A(13)$ and $A(18)$, require by far the smallest energy input ($0.06$ and $0.04$~eV per primitive cell, corresponding to $\Delta T_\mathrm{eff}\approx11$ and $7$~K), both far below the CDW transition temperature $T_\mathrm{CDW}=263$~K; the low-frequency mode $A(30)$ likewise remains readily accessible ($\Delta T_\mathrm{eff}\approx27$~K). In contrast, the remaining Se-dominated modes require $0.50$--$6.24$~eV per cell, corresponding to $\Delta T_\mathrm{eff}$ from $\sim90$~K up to $\sim1100$~K, i.e., comparable to or far exceeding $T_\mathrm{CDW}$. This reproduces the central energetic hierarchy identified with PBE and confirms that the experimental feasibility of the low-frequency CDW-amplitude pathway is not an artifact of the exchange--correlation functional.

\begin{table}[htbp]
\caption{Zone-center Raman-active $A$ modes of (TaSe$_4$)$_2$I in the CDW phase calculated with the PBEsol functional. Listed are the mode index, vibrational frequency (Freq), the mode-dependent normal-mode amplitude ($Q_\mathrm{c}$) at which the direct gap along $\Gamma\text{--}\mathrm{Z}$ is reduced to its minimum value $E_G^{\min}$, and the corresponding atomic displacement ($D_\mathrm{c}$). The last two columns give the harmonic coherent-mode energy at $Q_\mathrm{c}$, $E(Q_\mathrm{c})=\frac{1}{2}(2\pi f)^2 Q_\mathrm{c}^2$, evaluated per 22-atom primitive CDW cell, and the corresponding effective temperature rise $\Delta T_\mathrm{eff}=E(Q_\mathrm{c})/(3N k_\mathrm{B})$ ($N=22$) assuming full thermalization of the deposited energy over all lattice degrees of freedom.}
\label{tab:S_pbesol_qc}
\centering
\begin{ruledtabular}
\begin{tabular}{c c c c c c c}
Idx & \makecell{Freq\\(THz)} &
\makecell{$Q_\mathrm{c}$\\(\AA$\sqrt{\mathrm{amu}}$)} &
\makecell{$D_\mathrm{c}$\\(\AA)} &
\makecell{$E_G^{\min}$\\(meV)} &
\makecell{$E(Q_\mathrm{c})$\\(eV)} &
\makecell{$\Delta T_\mathrm{eff}$\\(K)} \\
\hline
13 & 2.03 & 2.7 & 0.075 & 0.4  & 0.06 & 11   \\
18 & 2.24 & 1.9 & 0.062 & 0.5  & 0.04 & 7    \\
30 & 4.29 & 2.0 & 0.044 & 0.9  & 0.15 & 27   \\
32 & 4.39 & 5.6 & 0.194 & 0.7  & 1.24 & 218  \\
35 & 4.70 & 8.9 & 0.337 & 2.3  & 3.59 & 631  \\
40 & 4.93 & 5.0 & 0.188 & 2.2  & 1.25 & 219  \\
41 & 5.04 & 6.0 & 0.264 & 1.1  & 1.87 & 330  \\
44 & 5.20 & 3.0 & 0.137 & 1.9  & 0.50 & 88   \\
47 & 5.39 & 4.3 & 0.146 & 1.9  & 1.10 & 194  \\
50 & 5.58 & 3.8 & 0.148 & 1.7  & 0.92 & 162  \\
59 & 7.77 & 7.1 & 0.248 & 50.5 & 6.24 & 1097 \\
63 & 7.88 & 5.7 & 0.204 & 3.4  & 4.14 & 727  \\
\end{tabular}
\end{ruledtabular}
\end{table}

Among the PBEsol modes, \(A(13)\) and \(A(18)\) are both closely related to the CDW-amplitude distortion. Their Ta-sublattice components follow the same tetramerization-modulating pattern discussed for the main-text \(A(18)\) mode, namely the motion that weakens or reverses the long--short Ta--Ta bond disproportionation along the quasi-one-dimensional chains. However, the accompanying Se displacements differ between these two modes, indicating that the CDW-amplitude character is split between two symmetry-preserving Raman modes in the PBEsol phonon spectrum. Therefore, these PBEsol distortions should not be interpreted as a simple rigid structural path back to the high-temperature non-CDW phase. In particular, although they strongly suppress the CDW gap and generate Weyl nodes, the resulting Weyl nodes occur at generic \(k\) points and do not exactly reproduce the Weyl-point configuration of the non-CDW structure.

\begin{table}[htbp]
\caption{Zone-center phonon modes of (TaSe$_4$)$_2$I in the CDW phase calculated using the PBEsol functional, including vibrational frequencies (THz), irreducible representations (Irrep), optical activity, and the space group retained after displacing atoms along each phonon eigenvector (SG$_\mathrm{disp}$).}
\label{tab:S_pbesol_phonons}
\centering
\scriptsize
\setlength{\tabcolsep}{2.8pt}
\begin{ruledtabular}
\begin{tabular}{r r c l l | r r c l l | r r c l l}
Idx & Freq & Irrep & Activity & SG$_\mathrm{disp}$ &
Idx & Freq & Irrep & Activity & SG$_\mathrm{disp}$ &
Idx & Freq & Irrep & Activity & SG$_\mathrm{disp}$ \\
\hline
1  & 0.00 & \(B_1\)    & Acoustic & $F$222 & 23 & 3.21 & \(B_1\) & IR,Raman & $C$2   & 45 & 5.33 & \(B_2\) & IR,Raman & $C$2 \\
2  & 0.00 & \(B_2\)    & Acoustic & $F$222 & 24 & 3.45 & \(B_2\) & IR,Raman & $C$2   & 46 & 5.37 & \(B_3\) & IR,Raman & $C$2 \\
3  & 0.00 & \(B_3\)    & Acoustic & $F$222 & 25 & 3.48 & \(B_1\) & IR,Raman & $C$2   & 47 & 5.39 & \(A\)   & Raman    & $F$222 \\
4  & 1.10 & \(B_3\) & IR,Raman & $C$2   & 26 & 3.89 & \(B_3\) & IR,Raman & $C$2   & 48 & 5.47 & \(B_2\) & IR,Raman & $C$2 \\
5  & 1.35 & \(B_3\) & IR,Raman & $C$2   & 27 & 4.17 & \(B_2\) & IR,Raman & $C$2   & 49 & 5.47 & \(B_1\) & IR,Raman & $C$2 \\
6  & 1.36 & \(B_1\) & IR,Raman & $C$2   & 28 & 4.19 & \(B_1\) & IR,Raman & $C$2   & 50 & 5.58 & \(A\)   & Raman    & $F$222 \\
7  & 1.40 & \(B_2\) & IR,Raman & $C$2   & 29 & 4.23 & \(B_3\) & IR,Raman & $C$2   & 51 & 5.75 & \(B_3\) & IR,Raman & $C$2 \\
8  & 1.72 & \(B_2\) & IR,Raman & $C$2   & 30 & 4.29 & \(A\)   & Raman    & $F$222 & 52 & 5.82 & \(B_2\) & IR,Raman & $C$2 \\
9  & 1.76 & \(B_1\) & IR,Raman & $C$2   & 31 & 4.37 & \(B_1\) & IR,Raman & $C$2   & 53 & 6.06 & \(B_1\) & IR,Raman & $C$2 \\
10 & 1.87 & \(B_2\) & IR,Raman & $C$2   & 32 & 4.39 & \(A\)   & Raman    & $F$222 & 54 & 6.06 & \(B_3\) & IR,Raman & $C$2 \\
11 & 1.88 & \(B_1\) & IR,Raman & $C$2   & 33 & 4.41 & \(B_2\) & IR,Raman & $C$2   & 55 & 6.34 & \(B_2\) & IR,Raman & $C$2 \\
12 & 1.89 & \(A\)   & Raman    & $F$222 & 34 & 4.62 & \(B_3\) & IR,Raman & $C$2   & 56 & 6.45 & \(B_1\) & IR,Raman & $C$2 \\
13 & 2.03 & \(A\)   & Raman    & $F$222 & 35 & 4.70 & \(A\)   & Raman    & $F$222 & 57 & 6.87 & \(B_1\) & IR,Raman & $C$2 \\
14 & 2.12 & \(B_1\) & IR,Raman & $C$2   & 36 & 4.72 & \(B_2\) & IR,Raman & $C$2   & 58 & 6.97 & \(B_2\) & IR,Raman & $C$2 \\
15 & 2.13 & \(B_2\) & IR,Raman & $C$2   & 37 & 4.73 & \(B_1\) & IR,Raman & $C$2   & 59 & 7.77 & \(A\)   & Raman    & $F$222 \\
16 & 2.14 & \(B_3\) & IR,Raman & $C$2   & 38 & 4.77 & \(B_3\) & IR,Raman & $C$2   & 60 & 7.83 & \(B_2\) & IR,Raman & $C$2 \\
17 & 2.20 & \(B_3\) & IR,Raman & $C$2   & 39 & 4.91 & \(B_3\) & IR,Raman & $C$2   & 61 & 7.86 & \(B_3\) & IR,Raman & $C$2 \\
18 & 2.24 & \(A\)   & Raman    & $F$222 & 40 & 4.93 & \(A\)   & Raman    & $F$222 & 62 & 7.88 & \(B_1\) & IR,Raman & $C$2 \\
19 & 2.93 & \(B_1\) & IR,Raman & $C$2   & 41 & 5.04 & \(A\)   & Raman    & $F$222 & 63 & 7.88 & \(A\)   & Raman    & $F$222 \\
20 & 2.98 & \(B_2\) & IR,Raman & $C$2   & 42 & 5.10 & \(B_3\) & IR,Raman & $C$2   & 64 & 8.06 & \(B_3\) & IR,Raman & $C$2 \\
21 & 3.07 & \(A\)   & Raman    & $F$222 & 43 & 5.18 & \(B_1\) & IR,Raman & $C$2   & 65 & 8.24 & \(B_2\) & IR,Raman & $C$2 \\
22 & 3.18 & \(B_2\) & IR,Raman & $C$2   & 44 & 5.20 & \(A\)   & Raman    & $F$222 & 66 & 8.27 & \(B_1\) & IR,Raman & $C$2 \\
\end{tabular}
\end{ruledtabular}
\end{table}

%\clearpage
%\renewcommand{\bibsection}{\section*{References}}
%\bibliographystyle{naturemag}
%\bibliography{refabbrev, ref}

\end{document}